\documentclass[aps,twocolumn,superscriptaddress,natlib,showpacs]{revtex4-1}
\setcitestyle{authoryear,round}
\usepackage{graphicx}
\usepackage{dcolumn}
\usepackage{bm}
\usepackage{color}
\usepackage{amssymb}
\usepackage{amsmath}
\usepackage{amscd} 
\usepackage{amssymb}

\begin{document}




\title{Perfect anomalous transport of subdiffusive cargos by molecular motors in viscoelastic cytosol }

\author{Igor Goychuk}

\email{igoychuk@uni-potsdam.de}
\affiliation{Institute for Physics and Astronomy, University of Potsdam, 
Karl-Liebknecht-Str. 24/25, 14476 Potsdam-Golm, Germany}

\begin{abstract}
Multiple experiments show that various submicron particles such as magnetosomes, RNA messengers, viruses, and even much smaller nanoparticles such as globular proteins diffuse anomalously slow in viscoelastic cytosol of living cells. Hence, their sufficiently fast directional transport by molecular motors such as kinesins is crucial for the cell operation. It has been shown recently that the traditional flashing Brownian ratchet models of molecular motors are capable to describe both normal and anomalous transport of such subdiffusing cargos by molecular motors with a very high efficiency. This work  elucidates further an important role of mechanochemical coupling in such an anomalous transport. It shows a natural emergence of a perfect subdiffusive ratchet regime due to allosteric effects, where the random rotations of a ``catalytic wheel'' at the heart of the motor operation become perfectly synchronized with the random stepping of a heavily loaded  motor, so that only one ATP molecule is consumed on average at each motor step along microtubule. However, the number of rotations made by the catalytic engine and the traveling distance both scale sublinearly in time. Nevertheless, this anomalous transport can be very fast in absolute terms. 
\end{abstract}






\maketitle
\section{Introduction}
\label{}
Intracellular transport by molecular motors is crucial for a eukaryotic cell operation (\cite{Pollard08,PhillipsBook,Nelson}).
This is especially true in view of the recent discoveries (\cite{Luby}) that various nanoparticle probes (\cite{Saxton,Guigas}), as well as naturally occurring biological nanoparticles such as proteins (\cite{WeissBJ04,BanksBJ05,WeigelPNAS}), viruses  (\cite{Seisenberg}),  RNA messengers (\cite{Golding}), various endosomes and granulates (\cite{Tolic,Jeon11,Caspi,Bruno11,TabeiPNAS}), including artificial magnetosomes (\cite{Robert}), and also lipids (\cite{KnellerJCP11,JeonPRL12}) subdiffuse either in membrane or in cytosol of living cells. This means that the mean-square distance covered by such particles scales sublinearly in time, $\langle \delta r^2(t)\rangle \sim 2D_{\alpha} t^\alpha/\Gamma(1+\alpha)$, where $\alpha$ is a  power law exponent of subdiffusion, $0<\alpha<1$, $D_\alpha$ is subdiffusion coefficient (within an effective 1d description), and $\Gamma(z)$ is familiar gamma-function. For example, magnetosomes of radius about $R=300$ nm subdiffuse in intact cytosol of PC3 tumor cells with $\alpha\approx 0.4$, and $D_\alpha\approx 
170\;{\rm nm^2/s^{0.4}}$, see in \cite{Robert,PCCP14,PhysBio15}. To subdiffuse over the distance of $2R$, such an endosome would require about $2.705\times 10^7$ seconds or about 313 days. Clearly a passive transport of such particles by subdiffusion on any significant distance within the cell is just impossible on  any physiologically relevant time scale. However, some cells must solve the tasks such as e.g. delivery of ion channels in a transfer bag provided by an endosome on the distances which can be even meter long, as e.g. in axons of some neuronal cells (\cite{Hirokawa}). So, how can cells solve such tasks even using an active transport by such molecular motors as kinesins, if cytosol is a gel-like viscoelastic medium causing subdiffusion? In particular, can such a transport be normal, rather than anomalously slow, in the sense that the traveling distance along the cell's microtubuli highways scales not sublinearly in time, $\langle \delta r(t)\rangle \sim  t^{\alpha_{\rm eff}}$, with some $\alpha\leq \alpha_{\rm eff}<1$, what is expected, but simply linearly with $\alpha_{\rm eff}=1$. Can such a transport be effective and sufficiently fast? And how? These are some challenging questions to be answered. 

\begin{figure}[ht]
  \centering
  \includegraphics[width=7cm]{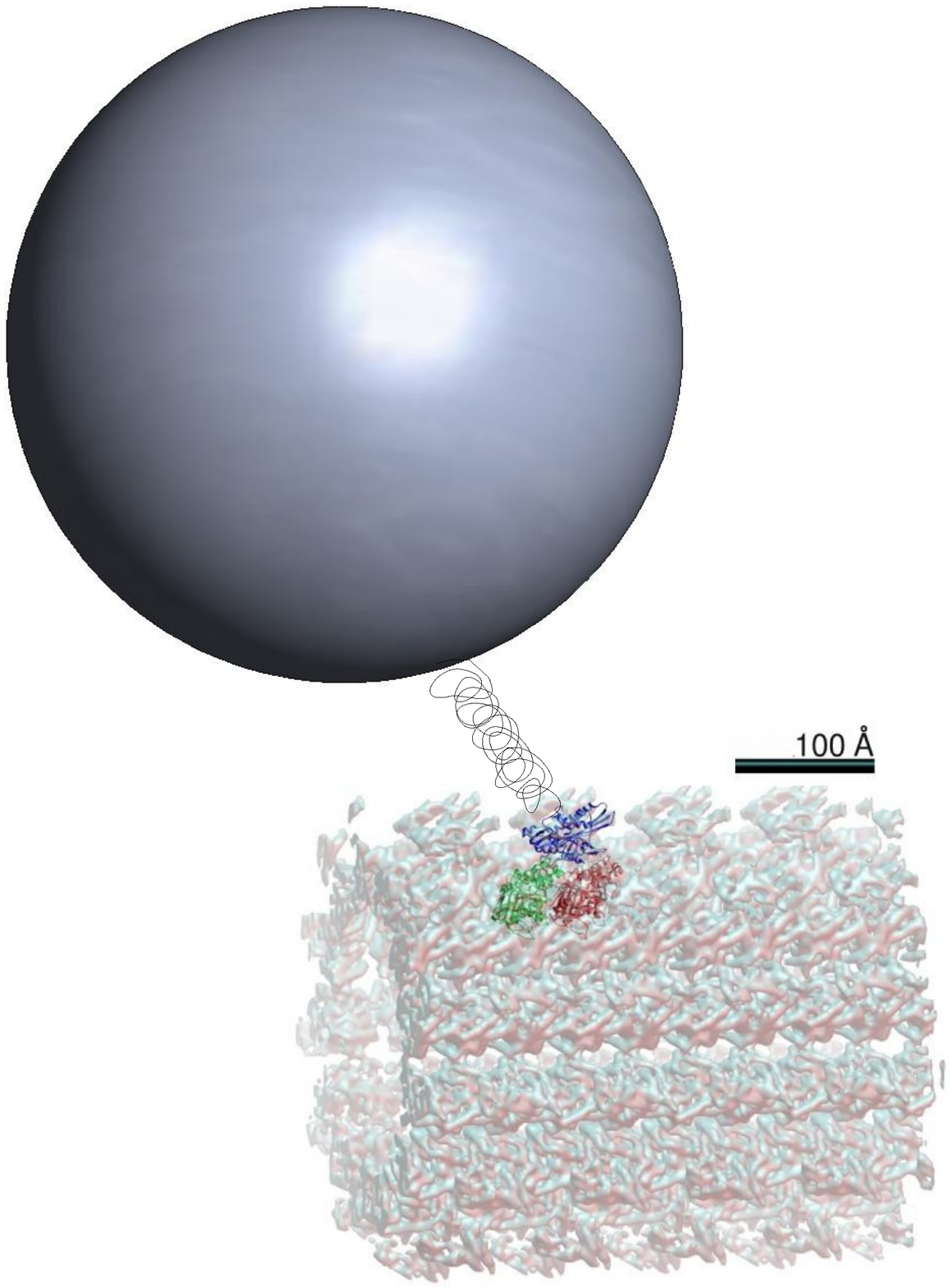}
 \caption{Kinesin walking on microtubule and pulling cargo on an elastic tether.}
  \label{Fig1}
\end{figure}

\begin{figure}[ht]
  \centering
  \includegraphics[width=9cm]{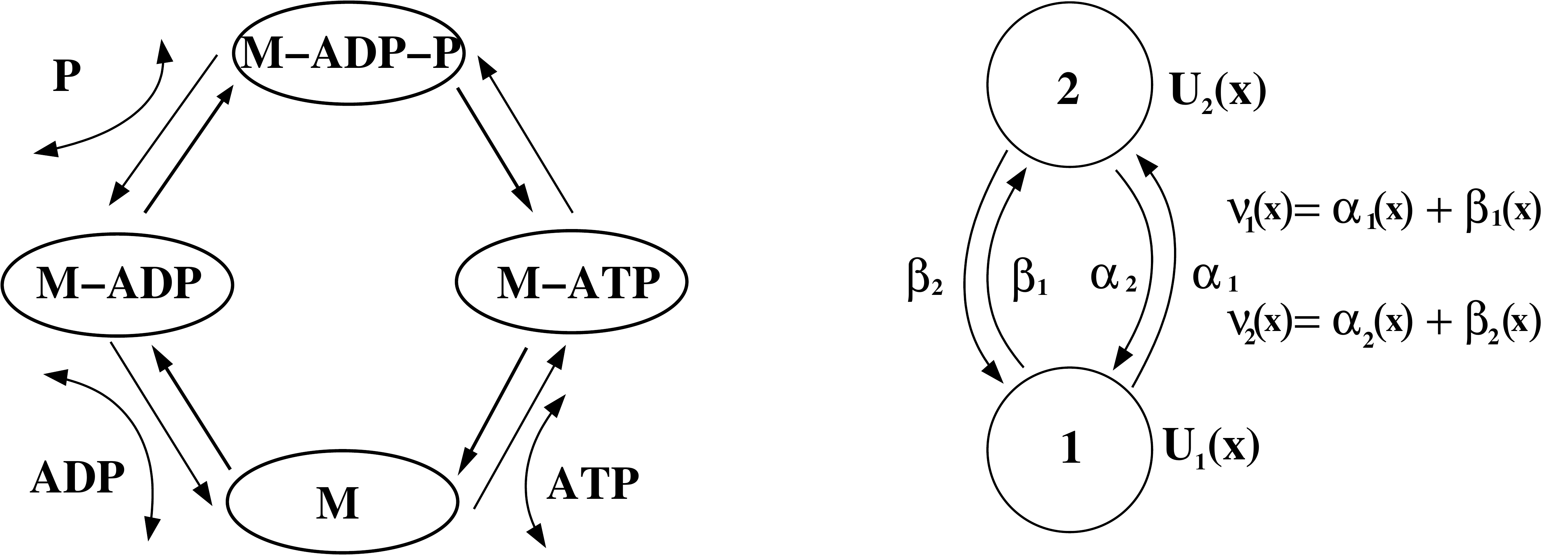}
 \caption{More realistic biochemical cycle for one motor head (left) and the minimal two-state model of cycling (right) that includes a binding potential change due to a change of the charge state of the motor. Spatial dependence of the transition rates on the coordinate along a periodic polar background of microtubule provides an allosteric mechanism of the mechano-chemical coupling. }
  \label{Fig2}
\end{figure}

The simplest modeling of a molecular motor is to represent it by a constant pulling force $F$ acting on a cargo, which,
otherwise, thermally subdiffuses (\cite{Caspi}), when it is not coupled to the motor. When the motor walks on microtubuli constituting a random transport network (kinesins), or it changes
its walking direction at random along the same track (myosins), the motor action on its cargo can be modeled by a fluctuating, non-thermal random
force $F(t)$ (\cite{Caspi,BrunoPRE}). In the absence of motors, thermal subdiffusion in viscoelastic media is
described   by a generalized Langevin equation or GLE (\cite{MasonPRL,AmblardPRL,Waigh}), with a memory friction
and thermal random force obeying the thermal fluctuation dissipation theorem (FDT), \cite{Kubo66,Zwanzig}. An algebraically slow memory
decay $\propto t^{-\alpha}$  yields subdiffusion $\langle \delta r^2(t)\rangle \sim  t^{\alpha_{\rm eff}}$, with
$\alpha_{\rm eff}=\alpha$, in the absence of non-thermal $F(t)$. The motor-driven diffusion has another exponent
$\beta$, $\langle \delta r^2(t)\rangle \propto  t^{\beta}$, with the maximal value $\beta_{\rm max}=2\alpha$
(\cite{BrunoPRE}) within this model. It can be superdiffusive only for $\alpha>0.5$. However, an experiment by \cite{Robert} in a
medium with e.g. $\alpha=0.4$ yielded $\beta=1.3\pm 0.1$ of active motor-assisted transport. It is essentially
larger than $2\alpha=0.8$. Also another experiment by \cite{Harrison} yielded $\beta\approx 1.74$ for the active transport with  $\alpha\approx 0.58$ of the passive transport.  Hence, such a modeling is far too simple and it cannot explain these experimental findings. A different
modeling route of flashing Brownian ratchets (\cite{AstumianBier,Julicher,Parmeggiani}) was taken by
\cite{PLoSONE14,PCCP14,PhysBio15,BeilsteinJ}. It is based on an extension of the previous research work on normal
diffusion Brownian ratchets, see e.g. review by \cite{ReimannReview}, onto the case of viscoelastic subdiffusion
featured by long-range memory correlations in the medium (\cite{GoychukPRE09,GoychukACP12}). Such kind of subdiffusion naturally
emerges in dense polymeric solutions, colloidal liquids and glasses, as well as cytosol of living cells
(\cite{Larson,Waigh,MasonPRL,AmblardPRL,GittesPRL,Santamaria,PanPRL,WeissPRE13}). A recent work by \cite{PCCP18} explains
how this kind of subdiffusion can win over the medium's disorder also featuring such complex heterogeneous media as
cytosol. 

Rocking ratchets of normal diffusion (\cite{MagnascoPRL,DoeringPRL,BartussekEPL}) have been generalized to viscoelastic subdiffusion by
\cite{ChemPhys10,PRE12a,MMNP13,PRE13}, and flashing ratchets (\cite{AidariFirst,ProstPRL,RousNature,AstumianPRL}) by \cite{NJP12}. The first application of flashing
subdiffusive ratchets to molecular motors pulling nanocargos was done by \cite{PLoSONE14}. In that work, motor and
cargo make one subdiffusing quasi-particle in assumptions that a tether between them is infinitely rigid, and a
spatially-asymmetric periodic ratchet potential acting on the motor stochastically switches between two realizations differing by a half of the spatial
period shift, like in \cite{Makhnovskii}. Moreover,  Markovian switching rates are identical and constant. Two such subsequent switches make one random cycle. The
mechano-chemical coupling is neglected in that earlier model. Depending on the size of cargo determining the subdiffusion
coefficient of combined quasi-particle, frequency of the binding potential flashing, loading
force applied, and other parameters, both anomalous and  normal transport regimes can be realized, \cite{PLoSONE14}. Very important is that
the ratchet transport of a subdiffusive cargo can be perfectly normal, in the sense that each switching cycle
results  on average in the transport step on a distance of the spatial period, and the averaged
number of such switches grows linearly in time (a perfect normal ratchet). However, anomalous transport regimes
can also be readily enforced. This model provides a principal framework to explain the origin of $\beta=1.3\pm 0.1$ for
$\alpha=0.4$ in \cite{Robert}, and also the origin  of $\beta\approx 1.74$ in \cite{Harrison},
where the passive subdiffusion of lipid droplets with $\alpha\approx 0.58$ is changed
to superdiffusion induced and assisted by
molecular motors. 

The tether or linker between the motor and its cargo is, however, never
infinitely rigid and the motor walking on microtubule is not fully exposed to viscoelastic constituents of cytosol.
For this reason, \cite{PCCP14} considered a more involved model with the motor being normally diffusing in a
ratchet potential of a similar kind (although, a different, saw-tooth form of the binding ratchet potential has
been chosen), whereas the cargo is subdiffusing in viscoelastic cytosol, and both particles are connected by
some elastic linker, like in Fig. \ref{Fig1}. Major earlier results were
confirmed within this more realistic model, which still lacked, however, a mechano-chemical coupling  between the
mechanical motion of the motor and cargo and the biochemical cycling of the motor in its intrinsic conformational
space. This drawback has been overcome by \cite{PhysBio15} who considered a very similar, in general features,
model for the motor as one by \cite{AstumianBier,Julicher,Parmeggiani}. It takes the mechano-chemical coupling
into account, and also the fact that any tether must have a finite maximal extension length. The related nonlinear effects were shown to be important, \cite{PhysBio15}, for weak tethers like one in \cite{Bruno11}.  This model is general and rich enough. It permits
different specific models for the mechano-chemical coupling, \cite{Parmeggiani}. One chosen by \cite{PhysBio15} (model A in this paper) allowed
to closely reproduce the earlier results in \cite{PCCP14} for the same amplitude of the ratchet potential,
$U_0=0.5\;{\rm eV}=20\; {k_BT_r}$. In this particular case, the mechano-chemical coupling is effectively absent,
and the motors perform cyclic turnovers with one almost fixed, position-independent rate for a very similar set of
parameters as in the earlier work, \cite{PCCP14}. However, already for $U_0=25\; {k_BT_r}$ and  $U_0=30\;
{k_BT_r}$ larger than the free energy of ATP hydrolysis, $\Delta G_{\rm ATP}=20\; {k_BT_r}$, used to drive one
biochemical cycle of the motor, the effects of mechano-chemical coupling become also very essential in the model
A. The most striking effect is that the number of motor turnovers and the number of ATP molecules hydrolyzed
during its operation start to scale sublinearly in time, $\langle N_{turn}(t)\rangle \propto t^{\gamma}$, with
$\alpha_{\rm eff}\leq \gamma \leq 1$. For the ratchet model with constant rates, $\gamma=1$ always. Since the
work against an external loading force scales as $t^{\alpha_{\rm eff}}$ and the energy consumed as $t^{\gamma}$, the
thermodynamic efficiency  generally decays in time as $1/t^\lambda$ with $\lambda=\gamma-\alpha_{\rm eff}$,
\cite{PhysBio15}. However, it can be appreciably large, over 50\% for a rather long time period (at the end of simulations corresponding to about 3 sec of physical time and traveling distances of the order of micrometer) at the maximum
sub-power of operation (\cite{PhysBio15,BeilsteinJ}). This regime requires, however, a large $U_0>\Delta G_{\rm ATP}$, with the stalling force about $10$ pN (for $U_0=30\;
{k_BT_r}$, \cite{PhysBio15}). It is essentially larger  than $5$ pN or $5.5-6.5$ pN observed for kinesins by \cite{SvobodaNature}, and \cite{Schnitzer00}, respectively.

The major question we address in this work is whether a similar regime is possible also for $U_0=\Delta G_{\rm ATP}=20\;k_BT_r$, and the stalling force in the range from 5 to 6 pN, as observed experimentally.
It will be shown that such a regime indeed emerges, however, for a different model  of mechano-chemical coupling (the model B below and in \cite{Parmeggiani}) such that it cannot be reduced to a ratchet model with constant switching rates in some range of parameters (like it happens within the model A). Moreover, the emergence of a perfect subdiffusive ratchet regime will be manifested with $\gamma=\alpha_{\rm eff}<1$, where thermodynamic efficiency does not decay in time. Such a perfect anomalous synchronization between anomalous biochemical turnovers of molecular motor and its mechanical motion due to a mechano-chemical coupling leads to a transport efficiency of nearly 
100\%, where consumption of one ATP molecule results into one step of the motor loaded with cargo along microtubule.


\section{Methods, Theory and Simulations}

We consider a model based on one studied earlier (\cite{AstumianBier,Julicher,Parmeggiani,PLoSONE14,PCCP14}). In essence, this is the same model as in \cite{PhysBio15}. Molecular motor moves in a flashing periodic saw-tooth ratchet potential, $U(x+L,\zeta(t))=U(x,\zeta(t))$, like one in the graphical abstract, with some potential height $U_0$. Here, $L=8$ nm is the spatial period of microtubule, \cite{Pollard08,PhillipsBook,SvobodaNature}, and $\zeta(t)$ is a conformational state of the motor.  Microtubuli are well-known to be polar, overally negatively 
charged periodic structures, \cite{Baker}, which provide transport highways for such motors as kinesins, \cite{Pollard08}. Hence, emergence of a periodic and asymmetric potential for charged nanoparticles, like molecular motor-proteins attached to microtubule, is quite natural. Furthermore, ATP molecules which serve as the source of free energy for the motors like kinesins or myosins, are also (negatively) charged, like are the products of the ATP hydrolysis: ADP and the phosphate group $P_i$. Thus, it is very natural that the binding potential flashes upon the conformational change of the motor related to its charge state fluctuations. The biochemistry of kinesin operation is very complex as it has two heads, with a simplest biochemical cycle depicted in the left part of Fig. \ref{Fig2}. The simplest theoretical model for its cycling is given in the right part of Fig. \ref{Fig2} (\cite{Hill,AstumianBier,Julicher}). This is a biochemical two-cycle or bi-cycle, with some four lump  rates.
 Of course, it presents a gross over-simplification, and hence, a truly minimal theoretical model. These rates are spatially-dependent, which expresses a mechano-chemical coupling, see below. In the spirit of this two-state model, one considers only two conformations, $\zeta_1$ and $\zeta_2$, with $\zeta(t)$ undergoing two-state fluctuations with spatially-dependent rates. Since two subsequent flashes make one cycle with the potential shifted by one spatial period, and the both motor heads are identical, it is natural to impose $U(x+L/2,\zeta_1)=U(x,\zeta_2)$ as an additional symmetry condition within this minimal model. Likewise, not only $\alpha_{1,2}(x+L)=\alpha_{1,2}(x)$, $\beta_{1,2}(x+L)=\beta_{1,2}(x)$, but also $\alpha_{1,2}(x+L/2)=\beta_{2,1}(x)$, etc. in this model. Furthermore, the energy $\Delta G_{\rm ATP}$ is used to rotate the ``catalytic wheel'' (\cite{Wyman,Rozenbaum,Qian}) in one preferred (counter-clockwise in Fig. \ref{Fig2}) direction. Thermodynamically this implies (\cite{Hill,Qian}) 
 \begin{eqnarray}
\frac{\alpha_1(x)\beta_2(x)}{\alpha_2(x)\beta_1(x)}=\exp[\Delta G_{\rm ATP}/(k_BT)],
\end{eqnarray}
for any $x$,  what can be satisfied, e.g.,  by choosing
\begin{eqnarray}
\frac{\alpha_1(x)}{\alpha_2(x)}& = & \exp[(U_1(x)-U_2(x)+\Delta G_{\rm ATP}/2)/(k_BT)], \nonumber \\
\frac{\beta_1(x)}{\beta_2(x)}& = & \exp[(U_1(x)-U_2(x)-\Delta G_{\rm ATP}/2)/(k_BT)] .
\end{eqnarray}
Furthermore, the total rates 
\begin{eqnarray}
\nu_1(x)&= &\alpha_1(x)+\beta_1(x), \nonumber \\
\nu_2(x)&= &\alpha_2(x)+\beta_2(x)
\end{eqnarray}
of the transitions between two energy profiles must satisfy
\begin{eqnarray}
\frac{\nu_1(x)}{\nu_2(x)}=\exp[(U_1(x)-U_2(x))/(k_BT)]
\end{eqnarray}
at thermal equilibrium. This is condition of the thermal detailed balance, 
where the dissipative fluxes vanish
both in the transport direction and within the conformational space of motor,
at the same time (\cite{Julicher,AstumianBier}). 
It is obviously satisfied 
for $\Delta G_{\rm ATP}\to 0$. There is still a lot of freedom in choosing rates, within the imposed requirements.
One possibility is to fix some $\alpha_1(x)=\beta_2(x+L/2)$.  Then, 
\begin{eqnarray}\label{nu1}
\nu_1(x) & = & \alpha_1(x+L/2)
\exp \left [-\frac{U_2(x)-U_1(x) + \Delta G_{\rm ATP}/2}{k_BT} \right ] \nonumber \\
& + & \alpha_1(x),  \\
 \nu_2(x) &= &  \alpha_1(x)\exp\left [-\frac{U_1(x)-U_2(x)+\Delta G_{\rm ATP}/2}{k_BT} \right]\nonumber \\ 
 &+& \alpha_1(x+L/2)\;. \label{nu2} \nonumber
 \end{eqnarray} 
 This is our model A.  Another choice is to fix $\alpha_2(x)=\beta_1(x+L/2)$. Then, 
\begin{eqnarray}\label{nuB1}
\nu_1(x) &=&  \alpha_2(x)  \exp\left [\frac{U_1(x)-U_2(x)
 + \Delta G_{\rm ATP}/2)}{k_BT}\right ] \nonumber \\
 &+& \alpha_2(x+L/2),\\
 \nu_2(x) &=&  \alpha_2(x+L/2)\exp \left [\frac{U_2(x)-U_1(x)+\Delta G_{\rm ATP}/2)}{k_BT}\right ] \nonumber \\ 
 &+& \alpha_2(x)\;. \label{nuB2} \nonumber 
 \end{eqnarray}  
 provides our model B, which is similar to the model B by \cite{Parmeggiani}. In both models, we shall assume that
 either $\alpha_1(x)=const$, or $\alpha_2(x)=const$, correspondingly, in a $\pm \delta/2$ neighborhood of the minimum of potential $U_1(x)$, and is zero otherwise. Using $\delta$ appropriately, $\delta=L/2$ in this paper, one can ensure that the enzyme turnovers can occur everywhere on microtubule, and not in some specially chosen domains only. 
 The difference between the models A and B seems subtle. However, the results are rather different, see below. In particular, the mechano-chemical coupling is markedly stronger in the model B. 
 
 The mechanical motion of the motor is mimicked by a Brownian particle subjected to the force
 $f(x,\zeta(t))=-\partial U(x,\zeta(t))/\partial x$ coming from the binding potential, viscous friction force $-\eta_m \dot x$, and  a thermal white Gaussian noise $\xi_m(t)$. The latter two are related by the second FDT, $\langle \xi_m(t)\xi_m(t') \rangle = 2k_BT \eta_m\delta(t-t')$ at the environmental temperature $T$.
 Inertial effects are neglected, like in the previous studies of molecular motors. Indeed, dynamics of nanoparticles in polymeric water solutions is typically overdamped. The inertial effects are present typically on the initial scales from picoseconds to nanoseconds, and we are interested in much longer times, up to seconds and minutes.
  Furthermore, the motor is assumed to be elastically coupled to a cargo (within a FENE model, \cite{FENE,PhysBio15}), with a spring constant $\kappa_L$ and a maximal extension length $r_{max}$. The limit $r_{max}\to \infty$ corresponds to a harmonic linker. Moreover, the motor is generally subjected also to a constant loading force $f_0$, which attempts to stop its directional motion being counter-directed. All in all, the motor is described by Eq. (\ref{model1a}) in 
 \begin{eqnarray}
\label{model1a}
\eta_m\dot{x} &=&
f(x,\zeta(t))-f_0 +\xi_m(t)+\frac{\kappa_L(y-x)}{1-(y-x)^2/r_{\rm max}^2}, \\
\eta_c \dot y &= &- \int_{0}^t\eta_{\rm mem}(t-t')\dot{y}(t')dt'\\ &-&\frac{\kappa_L(y-x)}{1-(y-x)^2/r_{\rm max}^2} \label{model1b}
+\xi_c(t)+\xi_{\rm mem}(t)\;.\nonumber
\end{eqnarray}
$f(x,\zeta(t))$ is piece-wise constant within the model considered. With the maximum of $U(x)$ dividing the potential period in the ratio $1:p$, $p>1$, it takes negative value $f_{-}=-(p+1)U_0/L$ within the spatial interval $[0,L/(p+1) )$ and positive value $f_{+}=(p+1)U_0/(pL)$ within the larger interval $[L/(p+1),L)$. Here, $U(0)=U(L)=0$. If flashing is sufficiently slow, so that the particle has time to relax to the potential minimum after each potential flash, it will be pushed forward by $f_+$ to a new potential minimum after each flash. In this way, a perfect ratchet transport mechanism can be realized, if flashing is also not far too slow, so that the particle does not have enough time to escape to another potential minimum being thermally agitated. For a high potential barrier $U_0\gg k_BT$, such escapes occur, however, very infrequently.
With an increasing loading force $f_0$, the potential barrier diminishes and it vanishes at $f_{\rm st}=f_{+}$, which is the stalling force in the absence of thermal fluctuations at $T=0$.
It must be mentioned, however, in this respect that at physiological temperatures the stalling force depends strongly on temperature.
To obtain it, $U_0$ should be replaced with a free energy barrier $U_0\to F_0=U_0-TS$, with $S\approx 11.2 \;k_B$ for $\alpha_1=170\;s^{-1}$ within the model A, \cite{PCCP14,PhysBio15}. The entropic component is as large as $T_rS\approx 11.2\; k_BT_r$ at $T_r=290$ K. Hence, to get a realistic stalling force for kinesin from 5 to 6 pN, \cite{SvobodaNature}, or about $7$ pN (\cite{Kojima}) at room temperatures, $U_0$ should be about $20\;k_BT_r$, or somewhat larger. For $U_0<15\;k_BT_r$, the above simple estimate does not work, cf. Fig. 6 in \cite{PCCP14}, and the stalling force is far too small, as compared to the experimental values.

Also elastic coupling to the cargo will generally strongly affect the motor operation. The cargo motion is described by Eq. (\ref{model1b}). It is subjected both to the viscous friction with the friction coefficient $\eta_c$ reflecting about 80\% of water content in cytosol, and to a viscoelastic memory friction characterized by the memory kernel $\eta_{\rm mem}(t)$. These frictional terms are related to the corresponding components of the thermal noise of the environment by the Kubo's second FDT, named also the fluctuation-dissipation relation or FDR, \cite{Kubo66,Zwanzig,WeissBook},   
 $\langle \xi_c(t)\xi_c(t') \rangle = 2k_BT \eta_c\delta(t-t'),
\langle \xi_{\rm mem}(t)\xi_{\rm mem}(t') \rangle = k_BT \eta_{\rm mem}(|t-t'|)$. Viscoelasticity with a complex shear modulus $G^*(\omega)\propto (i\omega)^\alpha$ (\cite{MasonPRL,Waigh,Larson}) corresponds to a strictly sub-Ohmic memory kernel,
 $\eta_{\rm mem}(t)=\eta_{\alpha}t^{-\alpha}/\Gamma(1-\alpha)$, $0<\alpha<1$, \cite{WeissBook},
with fractional friction coefficient $\eta_{\alpha}$, \cite{GoychukPRE09,GoychukACP12}. The corresponding memory term can be abbreviated as $\eta_\alpha d^\alpha y/dt^\alpha$ using the notion of fractional Caputo derivative (\cite{Mainardi97,Mathai17}). Furthermore, the corresponding thermal noise $\xi_{\rm mem}(t)$ is fractional Gaussian noise (fGn). It is a time derivative of the fractional Brownian motion (fBm) by \cite{Kolmogorov,KolmogorovTrans,Mandelbrot68}.
When the cargo is uncoupled to the motor ($\kappa_L=0$), spread of its position variance is described by  
\begin{eqnarray}\label{exact}
\langle\delta y^2(t)\rangle=2D_ctE_{1-\alpha,2}\left(-[t/\tau_{\rm in}]^{1-\alpha}
\right),
\end{eqnarray}
see in \cite{PRE13}, where $E_{a,b}(z)=\sum_{n=0}^\infty z^n/\Gamma(an+b)$ is the generalized
Mittag-Leffler function (\cite{Mathai17}), and $D_c=k_BT/\eta_c$ is normal diffusion coefficient. Initially,  at $t\ll \tau_{\rm in}=(\eta_c/\eta_\alpha)^{1/(1-\alpha)}$ diffusion is normal, $\langle\delta y^2(t) \rangle\approx 2D_c t$
whereas at large times, $t\gg \tau_{\rm in}$, it is anomalously slow, 
$\langle\delta y^2(t) \rangle\approx 2D_\alpha t^\alpha/\Gamma(1+\alpha)$. Here, $D_\alpha=k_BT/\eta_\alpha$ is the fractional diffusion coefficient 
 whose value plays a key role in anomalous transport processes.

\subsection{Markovian embedding}

Seen realistically, any power-law memory kernel has a long-time memory cutoff. Assuming it being exponential,
$\eta_{\alpha}\to \eta_{\alpha}\exp(-\nu_h t)$, an effective friction coefficient 
$\eta_{\rm eff}=\int_0^{\infty}\eta_{\rm mem}(t)dt=\eta_\alpha \tau_{\rm max}^{1-\alpha}$ can be introduced with $\tau_{\rm max}=1/\nu_h$. For $t\gg \tau_{\rm max}$, diffusion will be again normal with the diffusion coefficient 
$D_{c, \rm eff}=k_BT/(\eta_c+\eta_{\rm eff})$. However, $\tau_{\rm max}$ can be very large, in the range from tens of seconds to hours, see e.g. Table I in \cite{PRE12b} (for a different model of memory cutoff). Furthermore, a short-time memory cutoff $\tau_{\rm min}=1/\nu_0$ must also always exist on physical grounds, in any realistic description of a condensed medium beyond the continuous medium approximation. Here, $\nu_0$ is related to a maximal frequency of the mediums oscillators coupled to the Brownian particle within a dynamical theory of Brownian motion, \cite{WeissBook}. Hence, it is natural to approximate a power-law-scaling memory kernel between two memory cutoffs by a sum of exponentials, \begin{eqnarray}\label{kernel2}
\eta_{\rm mem}(t)=\sum_{i=1}^N k_i \exp(-\nu_i t),
\end{eqnarray}
obeying fractal scaling $\nu_i=\nu_0/b^{i-1}$, $k_i =C_\alpha(b)\eta_\alpha\nu_i^\alpha/\Gamma(1-\alpha)\propto \nu_i^\alpha$, where $C_\alpha(b)$
is some constant, which depends on $\alpha$ and $b$, \cite{PalmerPRL85,Hughes,GoychukPRE09,GoychukACP12}
 Obviously, $\tau_{\rm max}=b^{N-1}/\nu_0$. Depending on $b$ and $\alpha$, the accuracy of approximation can be between 4\% ($b=10$, $\alpha=0.5$) and 0.01\% ($b=2$, $\alpha=0.5$), see in \cite{MMNP13}.
 In fact, it provides an almost optimal approximation to the power law dependence, which can be slightly improved further, as suggested by \cite{Bohud07}.
Upon the use of the Prony series expansion (\ref{kernel2}), the non-Markovian dynamics of cargo allows for a multi-dimensional Markovian embedding, \cite{GoychukACP12}, by introducing auxiliary Brownian quasi-particles mimicking viscoelastic modes of environment with coordinates $y_i$ and frictional coefficients $\eta_i=k_i/\nu_i$. It reads
\begin{eqnarray}
\label{embedding1}
\eta_c\dot{y}&=& -\frac{\kappa_L(y-x)}{1-(y-x)^2/r_{\rm max}^2}-\sum_{i=1}^{N}k_i(y-y_i) \\
&+& \sqrt{2\eta_c k_BT}\xi_0(t), \nonumber\\
\eta_i\dot{y_i}&=&k_i(y-y_i)+\sqrt{2\eta_ik_BT}\xi_i(t),
\label{embedding2}
\end{eqnarray}
where $\xi_i(t)$ are uncorrelated white Gaussian noises of
unit intensity, $\langle \xi_i(t')\xi_j(t)\rangle=\delta_{ij}\delta(t-t')$, which
are also uncorrelated with the white Gaussian noise sources $\xi_0(t)$ and $\xi_m(t)$. 
To have a complete equivalence with the stated GLE model in Eqs.~(\ref{model1a}), (\ref{model1b})
with memory kernel (\ref{kernel2}), the initial positions $y_i(0)$ are sampled from independent Gaussian
distributions centered around $y(0)$, $\langle y_i(0)\rangle=y(0)$ with variances
$\langle[y_i(0)-y(0)]^2\rangle=k_BT/k_i$, \cite{GoychukACP12}. 
To see this equivalence, one has to (i) rewrite (\ref{embedding2}) in terms of viscoelastic force $u_i=k_i(y_i-y)$, 
(ii) formally solve the resulting equation for $u_i(t)$ with $\dot y(t)$ and $\xi_i(t)$ considered formally as some time-dependent functions and (iii) substitute the result, which consists of a regular part corresponding to friction with an exponentially decaying memory and a noise, into Eq. (\ref{embedding1}). Each noise component depends on $u_i(0)$, and all noise components are mutually independent. Considering $u_i(0)$ as random Gaussian variables with $\langle u_i(0)\rangle=0$ and  $\langle u_i^2(0)\rangle=k_i k_BT$, one can show that the resulting noise $\xi_{\rm mem}(t)$ 
is indeed a wide sense stationary Gaussian stochastic process which satisfies FDR with the memory function (\ref{kernel2}), see in \cite{GoychukPRE09,GoychukACP12} for detail. The resulting $\xi_{\rm mem}(t)$ presents a sum of independent Ornstein-Uhlenbeck processes which approximates fGn between two memory cutoffs.
Langevin equations (\ref{model1a}), 
(\ref{embedding1}), (\ref{embedding2}) considered together with a time-inhomogeneous 
Markovian process $\zeta(t)$, which is fully defined by two rates $\nu_{1,2}(x(t))$, provide a stochastic-dynamical description of the studied model. It is used in numerics, as described in the Supplementary Material.

\begin{table} 
\begin{center}
\caption{Parameter sets}\label{Table1}\vspace{0.5cm}

\begin{tabular}{|p{1 cm}|p{1.0 cm}|p{0.8cm}| p{0.8cm}|p{0.8 cm}| p{0.8cm}|}
\hline
Set, Model & $D_{0.4}$, ${\rm nm^2/s^{0.4}}$ & $\alpha_2$, $\rm{s}^{-1}$ & $\alpha_1$, $\rm{s}^{-1}$ & $U_0$, $k_BT_{ r}$  
& $r_{\rm rmax}$, nm \\
\hline
$S_{1}, A$ &  171 &  & 170 & 20 & $\infty$ \\ 

$S_2$, A &  1710 &  & 170 & 20 & $\infty$ \\

$S_5$, A &  171 &  & 34 & 20 & $\infty$ \\

$S_7$, A &  171 &  & 170 & 25 & 80 \\
$S_8$, A&  171 &  & 170 & 30 & 80 \\
$S_9$, A &  1710 &  & 170 & 25 & 80 \\
$S_{10}$, A &  1710 &  & 170 & 30 & 80 \\
\hline
$S_{1}$, B &  171 & 170  & & 20 & $\infty$ \\ 

$S_2$, B &  1710 & 170  &  & 20 & $\infty$ \\

$S_5$, B &  1710 & 34  &  & 20 & $\infty$ \\

$S_6$, B &  171 & 17  &  & 20 & $\infty$ \\

$S_7$, B &  1710 & 17  &  & 20 & $\infty$ \\

\hline
\end{tabular}
\end{center}
\end{table} 
 

\subsection{Choice of parameters and the details of numerics}

Like in  \cite{PCCP14,PhysBio15}, we use $a_m=100$ nm for the effective radius of kinesin,
about 10 times larger than its linear geometrical size (without tether) in order to account for the
enhanced effective viscosity experienced by the motor  partially exposed to cytosol compared
to its value in water. The viscous friction coefficient is estimated from the Stokes formula
as $\eta_m=6\pi a_m\zeta_w$, where $\zeta_w=1\;{\rm mPa\cdot s}$ is water viscosity used in calculations.
Furthermore, the time is scaled in the units $\tau_m=L^2\eta_m/U_0^*$  with $U_0^*=10\;k_BT_r$. For the
above parameters, $\tau_m\approx 2.94\;\mu{\rm s}$. Distance is scaled in units
of $L$,  elastic coupling constants in units of $U_0^*/L^2\approx 0.64$ pN/nm, and forces
in units of $U_0^*/L\approx 5.12$ pN. $\nu_0=100$ ($3.4\cdot 10^7$ 1/s) was chosen 
which corresponds to $\tau_{\rm min}=29.4$ ns,  
and $\alpha$ was $\alpha=0.4$, as found experimentally in \cite{Robert,Bruno11}. Two cargo sizes were considered, large $a_c=300$ nm, which corresponds
to the magnetosome size in  \cite{Robert}, and a ten times smaller one, like in Fig. \ref{Fig1}. For larger cargo, we assume
that its effective Stokes friction $\eta_c=6\pi a_c\zeta_w$ is enhanced by the factor 
of $\eta_{\rm eff}/\eta_c=3\cdot 10^4$ in cytosol. A particular embedding 
with $b=10$ and $N=10$ was chosen in accordance with our previous studies.
With these parameters, $\tau_{\rm max}=10^9\tau_{\rm min}=29.4$ s
and fractional friction coefficient 
$\eta_{\alpha}= \eta_{\rm eff}\tau_{\rm max}^{\alpha-1}/r$ with $r\approx 0.93$, \cite{PCCP14}.
The corresponding subdiffusion coefficient is
$D_{0.4}=k_BT/\eta_{0.4}\sim 1.71 \cdot 10^{-16}\;{\rm m^2/s^{0.4}}=171 \;{\rm nm^2/s^{0.4}}$,
in a semi-quantitative agreement with the experimental results in \cite{Robert}.
Smaller cargo is characterized by $\eta_{\rm eff}/\eta_c=3\cdot 10^3$ yielding 
$D_{0.4}=1710 \;{\rm nm^2/s^{0.4}}$, ten times larger. Furthermore, within the model A we used two values
of the rate constant $\alpha_1$: $170\;\rm{s}^{-1}$ (fast) and $34\;\rm{s}^{-1}$ (slow),
in order to match approximately the enzyme turnover rates $\nu \sim \alpha_1/2$
in Ref. \cite{PCCP14}. Accordingly, we used mostly $U_0=20$ $k_BT_{ r}$ in simulations, however,
also two larger values of $U_0$ were used, see Table \ref{Table1}, in order to arrive
at the thermodynamic efficiencies larger than 50\%. 
Within the model B we used three values of $\alpha_2$, see in Table \ref{Table1}.
The elastic spring constant is fixed to $\kappa_L=0.32$
pN/nm in this paper. A similar value was found in experiment, \cite{Kojima}.  For the maximal extension of linker
we used $r_{\rm rmax}=80$ nm, \cite{Pollard08}, and also $r_{\rm rmax}=\infty$, which corresponds to harmonic
linker in \cite{PCCP14}. As it has been shown earlier in \cite{PhysBio15}, for a strong linker considered, the harmonic approximation is, in principle, sufficient. Hence, within the model B in this paper we used only it. However, for weak linkers anharmonic effects can be very essential (\cite{PhysBio15}). Such weak linkers are not considered in this paper. The studied set of parameters is shown in Table \ref{Table1}. 

To numerically integrate stochastic Langevin dynamics for a fixed
potential realization $U_{1,2}(x)$, we used stochastic Heun method, see in the Supplementary Material, 
implemented in parallel on NVIDIA Kepler graphical processors.
Stochastic switching between two potential realizations is simulated using
a well-known algorithm. Namely, if the motor is moving on $U_1(x)$ or $U_2(x)$ surface, 
at each integration  time step $\Delta t$ it can switch with the 
probability $\nu_{1}(x)\Delta t$ or $\nu_{2}(x)\Delta t$, correspondingly, to another state, or to
evolve further on the same potential surface. Here,  $\nu_{1,2}(x)$ are the rates corresponding to either model A, or model B, see above.
We used  $\Delta t=5\cdot 10^{-3}$ for the 
integration time step and $n=10^3$  for the ensemble averaging.
The maximal time range of integration was $10^6$, which corresponds to $2.94$ sec 
of motor operation. Notice that a further increase of $N$ does not influence results, whereas it exponentially increases $\eta_{\rm eff}$ of cargos. This means that $\eta_{\alpha}$ and $D_\alpha$ are the truly relevant parameters of fractional transport, and not $\eta_{\rm eff}$,. Furthermore,
$\Delta G_{\rm ATP}=20$ $k_BT_r$ was taken in all numerical simulations, and $T=T_r=290$ K, so that $k_BT_r\approx 25$ meV.

\subsection{Stochastic energetics}

Stochastic energetics can be considered following \cite{Julicher,PhysBio15}. The useful work done by motor against a loading force $f_0$ is $W_{\rm use}(t)=f_0 \langle \delta x(t)\rangle \propto t^{\alpha_{\rm eff}}$, whereas the input energy that it consumes scales as $E_{\rm in}(t)=\Delta G_{\rm ATP}\langle N_{\rm turn}(t)\rangle$, where $\langle N_{\rm turn}(t)\propto t^\gamma$ is the number of turnovers $1\to 2\to 1$. This yields for thermodynamic efficiency
\begin{eqnarray}\label{eff2}
R_{\rm th}(t)=\frac{W_{\rm use}(t)}{\Delta G_{\rm ATP}\langle N_{\rm turn}(t)\rangle}\propto 1/t^{\gamma-\alpha_{\rm eff}}.
\end{eqnarray}
This definition is, however, not quite precise because it assumes
that all the turnovers of the ``catalytic wheel'' occur with ATP hydrolysis. However, some turnovers occur backwards, with ATP synthesis, within this model. Since such backward turnovers occur seldomly for $\Delta G_{\rm ATP}=0.5$ eV, Eq. (\ref{eff2}) only slightly underestimates the proper efficiency, see in \cite{PhysBio15}.
To correctly calculate consumption
of ATP molecules one should count $p_1\Delta G_{\rm ATP}/2$, with 
$p_1=(\alpha_1-\beta_1)/(\alpha_1+\beta_1)$ for the transition $U_1\to U_2$, and 
$p_2\Delta G_{\rm ATP}/2$
with  $p_2=(\beta_2-\alpha_2)/(\alpha_2+\beta_2)$ for the transition $U_2\to U_1$. A corresponding modification of (\ref{eff2}) will be named the proper efficiency.

\section{Results}

\begin{table} 
\begin{center}
\caption{Parameters of the fit with Eq. (\ref{fit}) within 1\% error tolerance done with the Levenberg-Marquardt algorithm using XMGRACE software,  \cite{XMGRACE}, and the corresponding values of $R_{\rm th}^{(\rm max)}$, and $f_{\rm max}$.}
\label{Table2}\vspace{0.5cm}
\begin{tabular}{|p{1.0 cm}|p{0.7 cm}|p{0.7cm}| p{0.8cm}|p{1.0 cm}|p{0.8 cm}|p{0.8 cm}|}
\hline
Set, Model & $k$ & $f_{\rm st}$, pN & $\epsilon$ & $q$ & $R_{\rm th}^{(\rm max)}$ &  $f_{\rm max}$, pN  \\
\hline
$S_{1}, A$ &  0.389 & 6.00 & 1.270 & 0.923 &  0.103 & 3.06\\ 

$S_2$, A &  0.447 & 5.89 & 2.032 &   -1.251  & 0.237 & 4.15\\

$S_5$, A &  0.458 & 4.81 & 3.447 & 0.653 & 0.194 & 2.88 \\

$S_7$, A &  0.823 & 9.05 & 1.152 & 0.307 & 0.284 &  5.18\\
$S_8$, A&  0.968 & 9.98 & 0.761 & -40.44 & 0.707 &  8.50 \\

$S_9$, A & 0.872 & 9.04 & 7.140 & -0.111 & 0.580 & 6.92 \\

$S_{10}$, A & 0.973 & 10.0 & 4.780 & -23.24 & 0.828 & 9.16 \\
\hline
$S_{1}$, B &  0.387 & 6.30 & 4.966 & 0.993 & 0.103 &  3.12\\ 

$S_2$, B &  0.606 & 6.29  &  6.981 & 0.707 & 0.331 &  4.16\\

$S_5$, B &  0.525 & 5.41 & 6.877 & 0.587 & 0.300 & 3.71 \\

$S_6$, B &  0.484 & 5.05  & 4.209 & 0.790 & 0.200 & 2.93 \\

$S_7$, B &  0.489 & 5.05  & 6.889 & 0.596 & 0.280 & 3.46\\
\hline
\end{tabular}
\end{center}
\end{table}

\begin{figure}[ht]
  \centering
  \includegraphics[width=7cm]{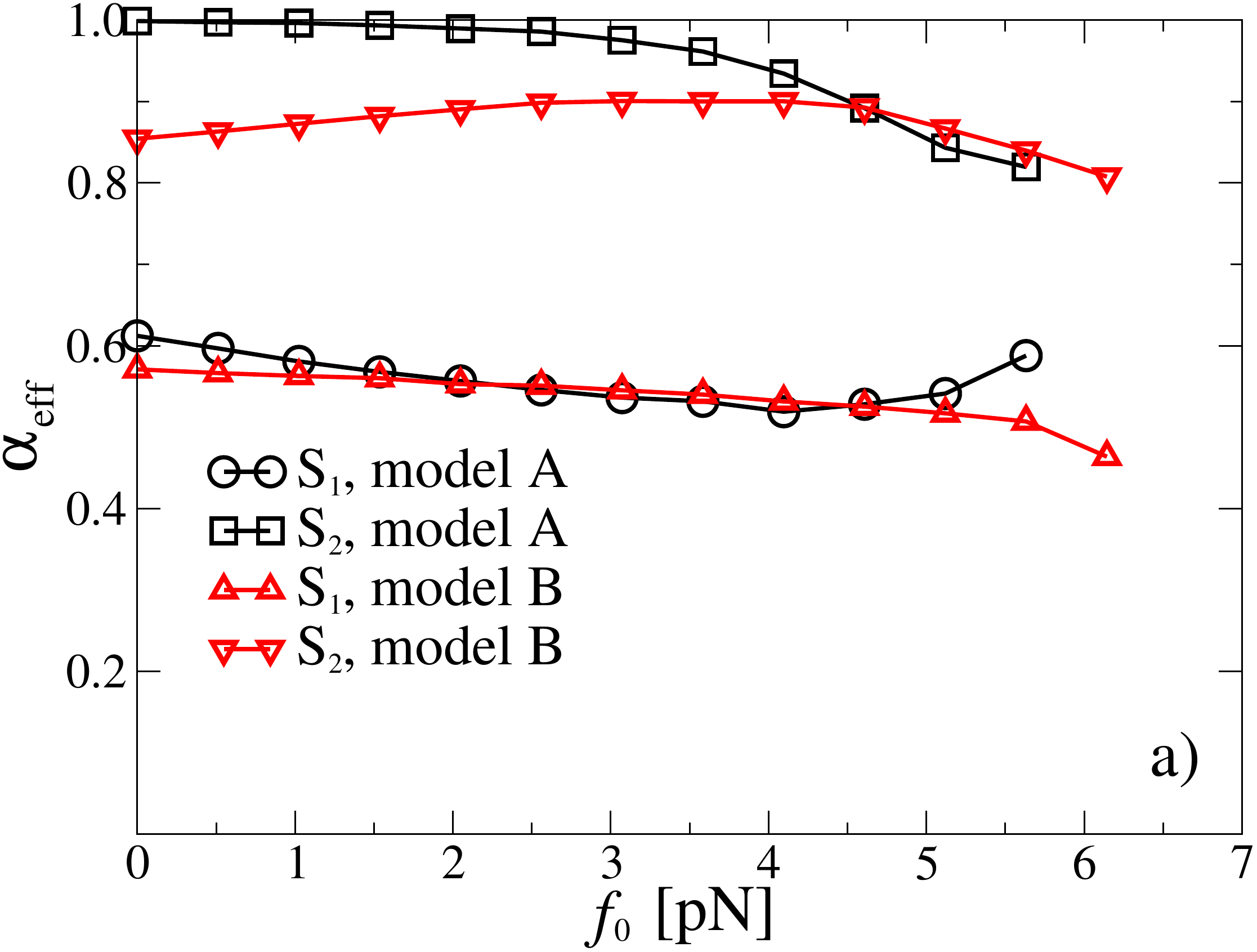}\\
  \includegraphics[width=7cm]{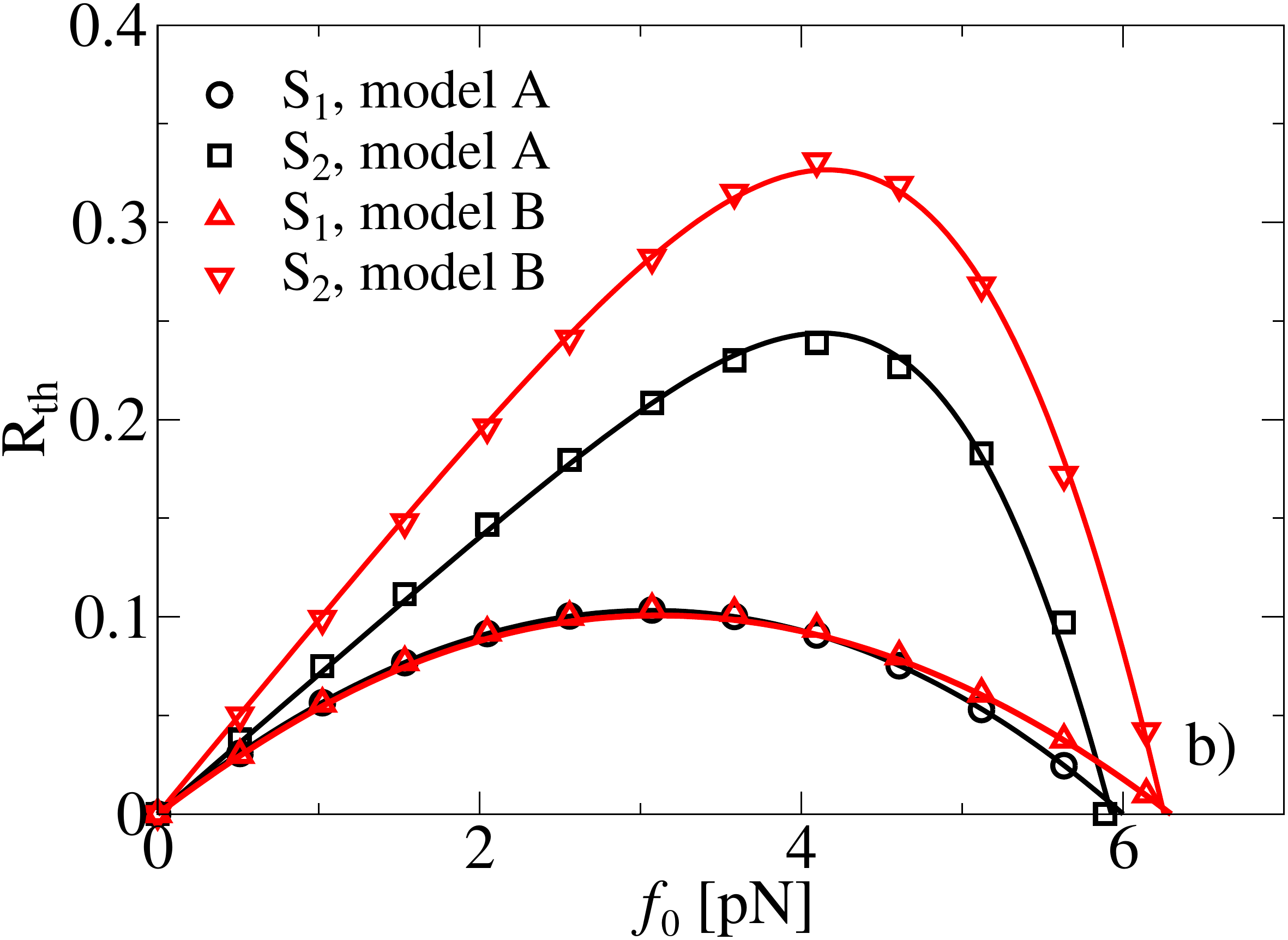} 
 \caption{(Color online). (a) Transport power exponent $\alpha_{\rm eff}$ and (b) thermodynamic efficiency for the sets $S_{1,2}$ in the models A and B \textit{vs.} loading force $f_0$. Notice that in the case $S_2$ the motor has an essentially larger efficiency within the model B than the model A, although the transport is more anomalous within the model B, especially for a small load. In part (b), full lines present fits with Eq. (\ref{fit}) with parameters shown in Table \ref{Table2}. Thermodynamic efficiency is calculated in accordance with Eq. (\ref{eff2}) at the end point of simulations. }
  \label{Fig3}
\end{figure}

We first made a comparative study of the dependence of the transport exponent $\alpha_{\rm eff}$ and thermodynamic efficiency on the loading force $f_0$ for two sets of parameters, $S_1$ and $S_2$, within the models A and B, see in  Fig. \ref{Fig3}. For the larger cargo, the set $S_1$, see in Table \ref{Table1}, the results do not differ much:
$\alpha_{\rm eff}$ is around $0.6$, which can explain $\beta=1.3\pm 0.1$ in \cite{Robert}. The maximal efficiency is about 10\% and the stalling force is slightly larger in the model B, $f_{\rm st}=6.30$ pN, \textit{vs.} $f_{\rm st}=6 $ pN in the model A, see in Table \ref{Table2}.
The numerical data on the efficiency are parametrized in this paper by the dependence,
\begin{eqnarray}\label{fit}
R_{\rm th}(f_0)=k\frac{f_0}{f_{\rm st}}\Bigg [ 1&&\\
&&-\frac{f_0/f_{\rm st}}{(1-q)\exp[\epsilon (1-f_0/f_{\rm st})]+q}\Bigg ],\nonumber
\end{eqnarray}
where $k$, $\epsilon$, $f_{\rm st}$, and $q$ are considered as fitting parameters with their values given in Table \ref{Table2}. It is derived from the assumptions that subvelocity $v_{\alpha}$, defined by $\langle \delta x(t)\rangle =v_{\alpha} t^{\alpha}/\Gamma(1+\alpha)$ with $\alpha=\alpha_{\rm eff}$ decays with $f_0$ as (see in Supplementary Material)
\begin{eqnarray}\label{fit_velo}
v_\alpha(f_0)=v_\alpha(0)\Bigg [ 1&&\\ &&-\frac{f_0/f_{\rm st}}{(1-q)\exp[\epsilon (1-f_0/f_{\rm st})]+q}\Bigg ],\nonumber
\end{eqnarray}
and $E_{\rm in}(t)$ does not depend on $f_0$. Here, $\epsilon$ is an energy barrier in the units of $k_BT$ and $q\leq 1$ is a parameter. For $q=1$, $v_\alpha(f_0)=v_\alpha(0)(1-f_0/f_{\rm st})$. Fit works actually pretty well with $q$ fixed to $q=0$. However, an almost perfect fit is obtained with an adjustable value of $q$.  Notice that $q$ can take negative values, which can be justified from diffusional models, \cite{GoychukPNAS}. This fit is not unique, see in the Supplementary Material for an alternative. However, it is biophysically better motivated. If these assumptions are justified, the maximum of $R_{\rm th}$ in Eq. (\ref{fit})
corresponds to thermodynamic efficiency at the maximum of sub-power. Whereas the latter assumption is well fulfilled within the model A at $U_0=20\; k_BT_r$, \cite{PhysBio15}, generally it is not correct, especially within the model B, see below, where the ATP consumption generally strongly depends on $f_0$. Hence, the maximum of the fit (\ref{fit}) not always correspond to the maximum at maximal sub-power, see below.  Nevertheless, it works nicely anyway. Its relation to the Jacobi efficiency in the linear operation regime of normal motors should also be mentioned. Indeed, it reduces to the Jacobi efficiency, see e. g. in \cite{BeilsteinJ}, at $\epsilon=0$ or $q=1$, $k=1$, and $\alpha_{\rm eff}=\gamma=1$.

\begin{table} 
\begin{center}
\caption{Power law exponents depending on the loading force $f_0$, the case $S_2$, model B. Fitting  values of $\alpha_{\rm eff}$ and $\gamma$ in this and other tables are obtained from the $\langle \delta x(t)\rangle$ and $L\langle N_{\rm turn}(t) \rangle$ dependencies, like ones presented in Fig. \ref{Fig5}, using a power law fit for the last 1 sec part of the trajectory. It is done using the Levenberg-Marquardt algorithm in XMGRACE software, \cite{XMGRACE}, with 1\% error tolerance, $\lambda=\gamma-\alpha_{\rm eff}$.}\label{Table3}\vspace{0.5cm}
\begin{tabular}{|p{1 cm}|p{1.5 cm}|p{1.5 cm}| p{1.5 cm}|}
\hline
$f_0$, pN & $\alpha_{\rm eff}$ & $\gamma$ &  $\lambda $\\
\hline
0 & 0.854231 & 0.856338  & 0.002107 \\ 
0.512 & 0.863078 & 0.865964 & 0.002886\\ 
1.024 & 0.872694 & 0.877308 & 0.004614 \\ 
1.536 & 0.882037 & 0.888624 & 0.006587\\ 
2.048 & 0.890468  & 0.900774 & 0.010306\\ 
2.560 & 0.898387  & 0.914309  & 0.015922\\ 
3.072 & 0.900635 & 0.926184 & 0.025549\\ 
3.584 & 0.900195  & 0.940803 & 0.040608\\ 
4.096 & 0.900237  & 0.964067 & 0.063830\\ 
4.608 & 0.892551 & 0.988937 & 0.096386\\ 
5.120 & 0.890468  & 1 & 0.109532\\ 
5.632 & 0.839736   &   1 & 0.160264 \\
6.148 & 0.807682 & 1 & 0.192318 \\
\hline
\end{tabular}
\end{center}
\end{table} 

\begin{table} 
\begin{center}
\caption{Power law exponents depending on the loading force $f_0$, the case $S_6$, model B. }\label{Table4}\vspace{0.5cm}
\begin{tabular}{|p{1 cm}|p{1.5 cm}|p{1.5 cm}| p{1.5 cm}|}
\hline
$f_0$, pN & $\alpha_{\rm eff}$ & $\gamma$ &  $\lambda $\\
\hline
0 & 0.778586  &  0.799655 & 0.021069 \\ 
0.512 & 0.792495 & 0.823397 & 0.030902\\ 
1.536 & 0.798144 & 0.868486 & 0.070342\\ 
2.048 & 0.799791  &  0.899897 & 0.100106\\ 
2.560 & 0.793568  &  0.937876 & 0.144308 \\ 
3.072 & 0.78149 & 0.973073 & 0.191583 \\ 
3.584 & 0.760739  & 1 & 0.239261\\ 
4.096 & 0.72811  & 1 & 0.27189\\ 
4.608 & 0.699885 & 1 & 0.300115\\ 
5.120 & 0.687050  & 1 &0.312950 \\ 
\hline
\end{tabular}
\end{center}
\end{table} 

\begin{table}[!ht] 
\begin{center}
\caption{Power law exponents depending on the loading force $f_0$, the case $S_5$, model A.}\label{Table5}\vspace{0.5cm}
\begin{tabular}{|p{1 cm}|p{1.5 cm}|p{1.5 cm}| p{1.5 cm}|}
\hline
$f_0$, pN & $\alpha_{\rm eff}$ & $\gamma$ &  $\lambda $\\
\hline
0 & 0.995152 & 1.00  & 0.004848\\ 
0.512 & 0.991188  & 1.00 & 0.008812\\ 
1.024 & 0.982147 & 1.00 & 0.017853 \\ 
1.536 & 0.969404 & 1.00 & 0.030596\\ 
2.048 &  0.925052 & 1.00 & 0.074948\\ 
2.560 & 0.909704  & 1.00  & 0.090296 \\ 
3.072 & 0.813889 & 1.00 & 0.186111\\ 
3.584 & 0.756685  & 1.00 & 0.243315\\ 
4.096 &  0.697763 & 1.00 & 0.302237\\ 
4.608 & 0.618159 & 1.00 & 0.381841\\ 
\hline
\end{tabular}
\end{center}
\end{table}
\begin{table}[!ht] 
\begin{center}
\caption{Power law exponents depending on the loading force $f_0$, the case $S_5$, model B.}\label{Table6}\vspace{0.5cm}
\begin{tabular}{|p{1 cm}|p{1.5 cm}|p{1.5 cm}| p{1.5 cm}|}
\hline
$f_0$, pN & $\alpha_{\rm eff}$ & $\gamma$ &  $\lambda $\\
\hline
0 & 0.928004  &  0.928294 & 0.000290 \\ 
0.512 & 0.937161 & 0.937699  & 0.000538\\ 
1.024 & 0.945399  & 0.946257 & 0.000858\\ 
1.536 & 0.950206 & 0.952371 & 0.002165\\ 
2.048 &  0.952820   & 0.956787 & 0.003967\\ 
2.560 & 0.956454  & 0.963371 & 0.006917\\ 
3.072 & 0.957644 & 0.970018 & 0.012374\\ 
3.584 & 0.956840  & 0.978205 & 0.021365\\ 
4.096 & 0.954589  & 0.989856 & 0.035267\\ 
4.608 & 0.938140  & 0.999875 & 0.061735\\ 
5.120 &  0.917003 & 1.00 & 0.082997\\ 
\hline
\end{tabular}
\end{center}
\end{table} 
 
\begin{figure}[ht]
  \centering
  \includegraphics[width=7cm]{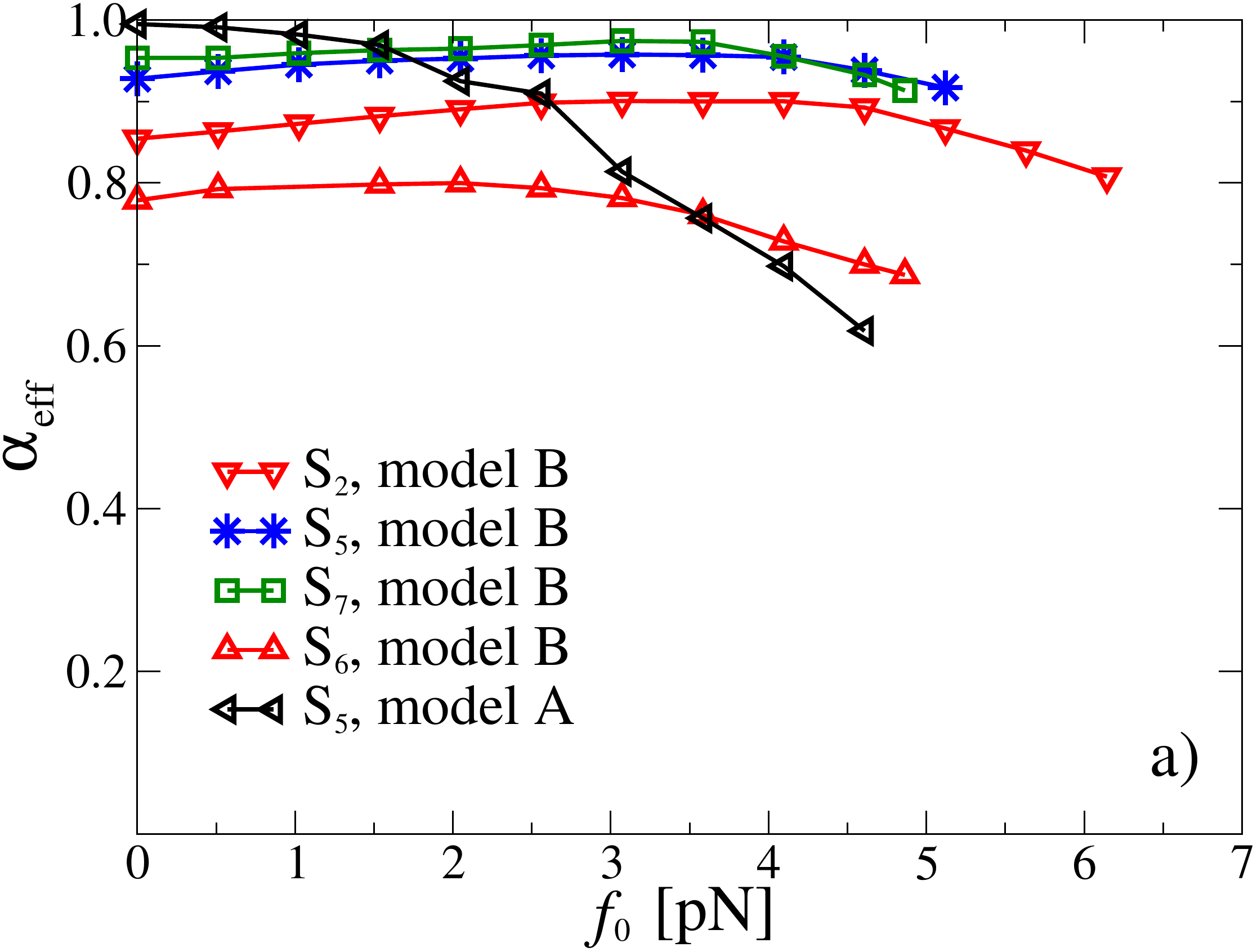}\\
  \includegraphics[width=7cm]{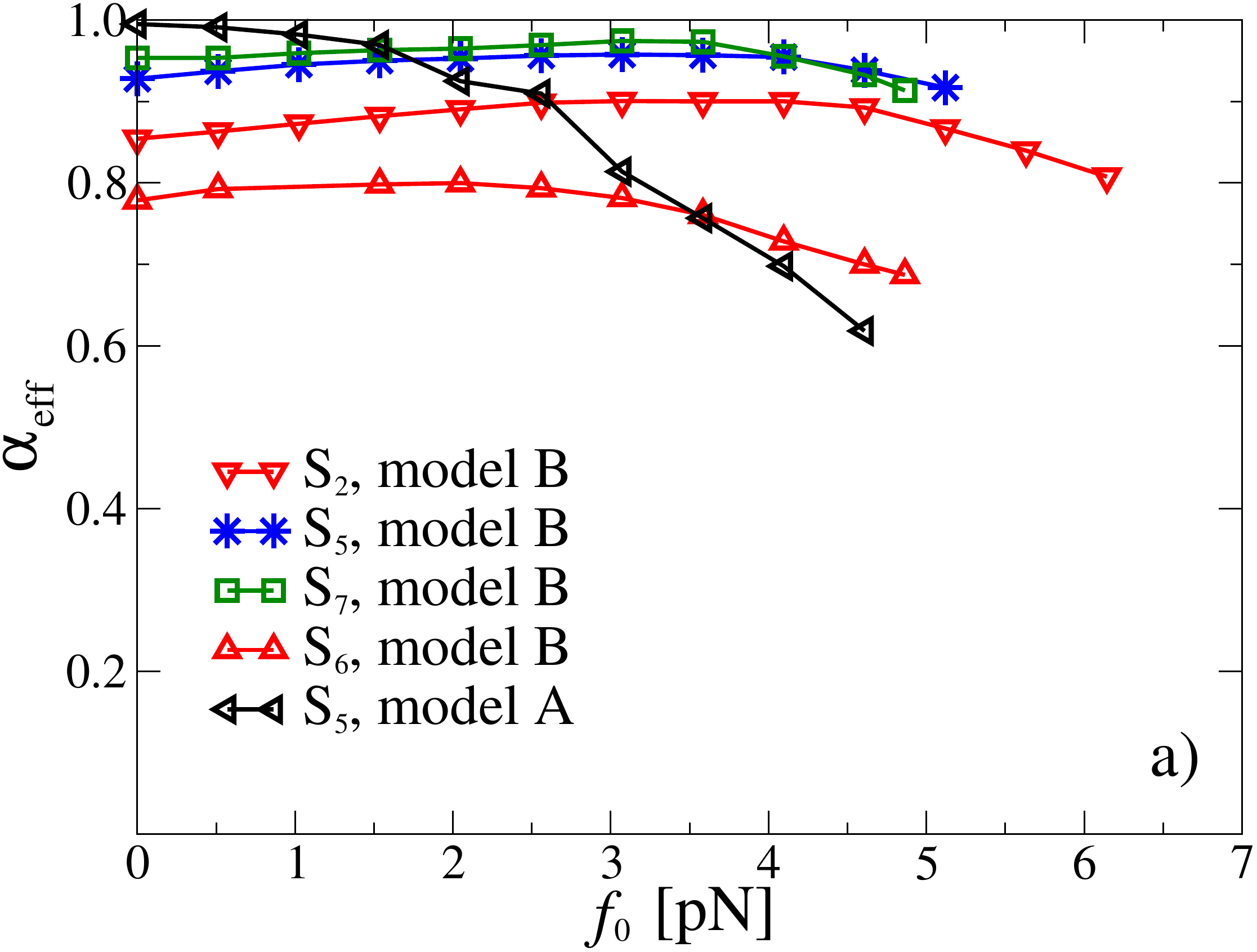} 
 \caption{(Color online). (a) Transport power exponent $\alpha_{\rm eff}$ and (b) thermodynamic efficiency \textit{vs.} loading force $f_0$ for 
 several other sets shown in the plots and discussed in the text. In part (b), full lines present fits with Eq. (\ref{fit}) and parameters shown in Table \ref{Table2}. Thermodynamic efficiency is calculated in accordance with Eq. (\ref{eff2}).}
  \label{Fig4}
\end{figure}

\begin{figure*}[!]
  \centering
  \includegraphics[width=7cm]{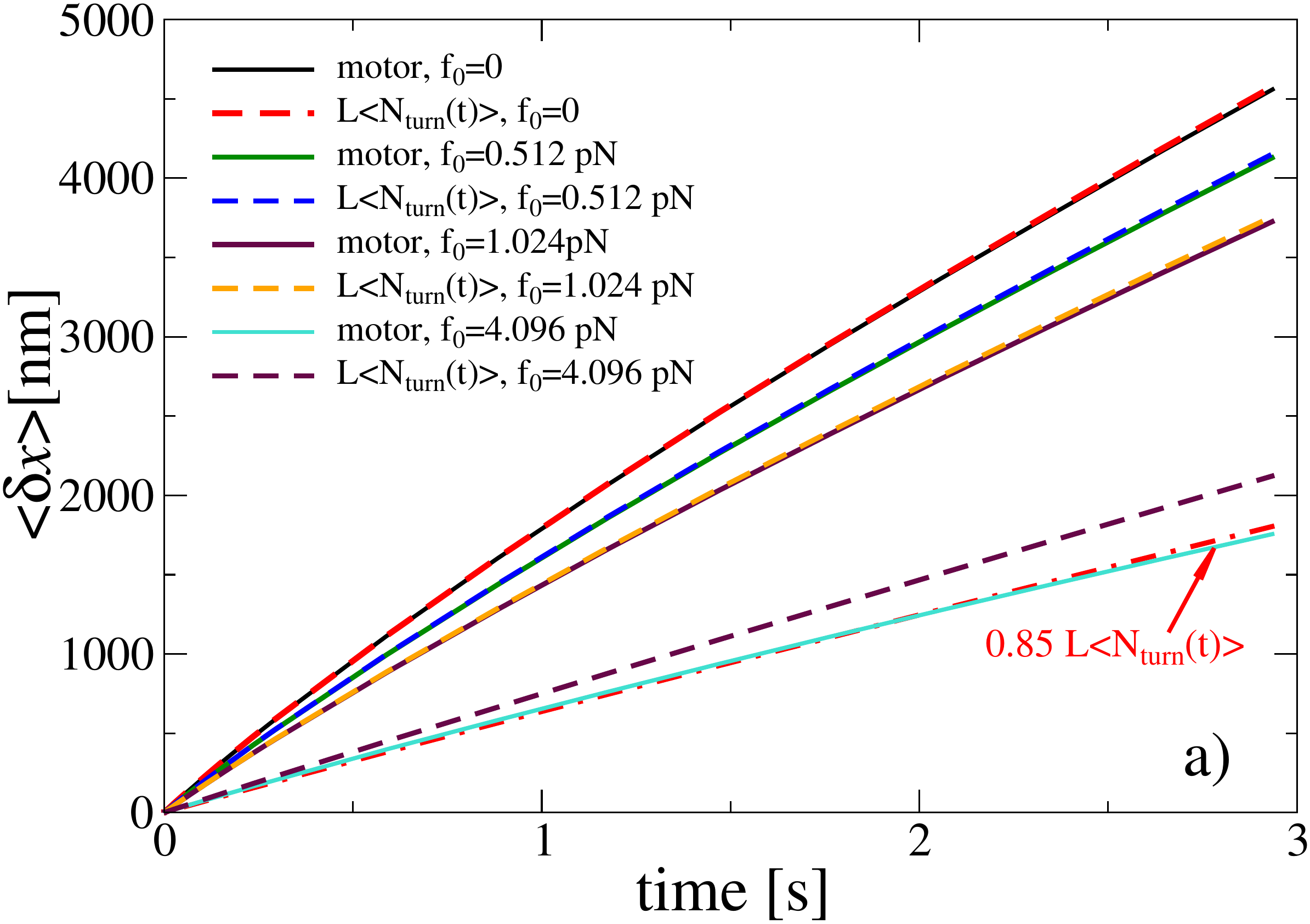}
  \hspace{0.8cm}
  \includegraphics[width=7cm]{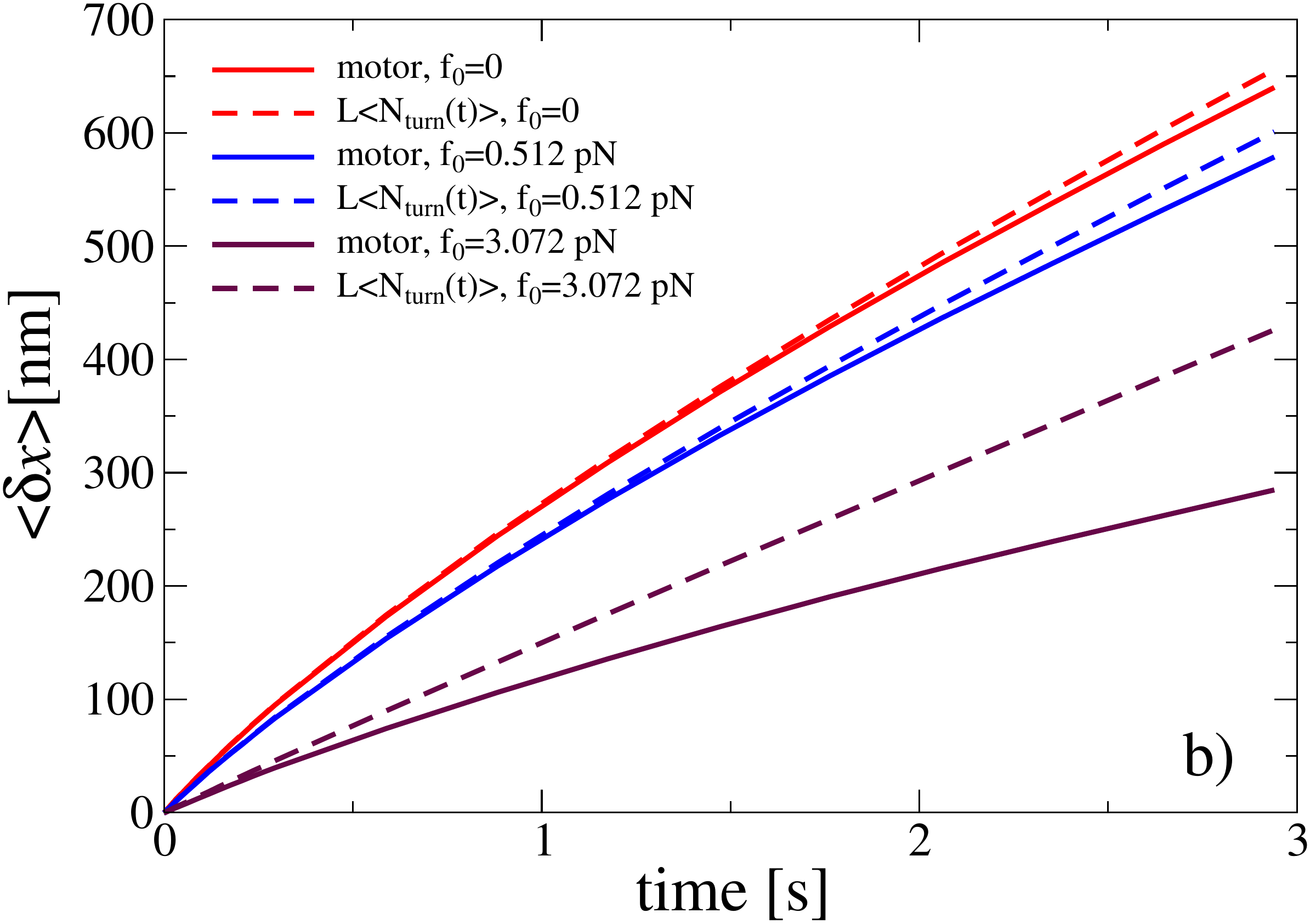} \\ \vspace{0.8cm}
\includegraphics[width=7cm]{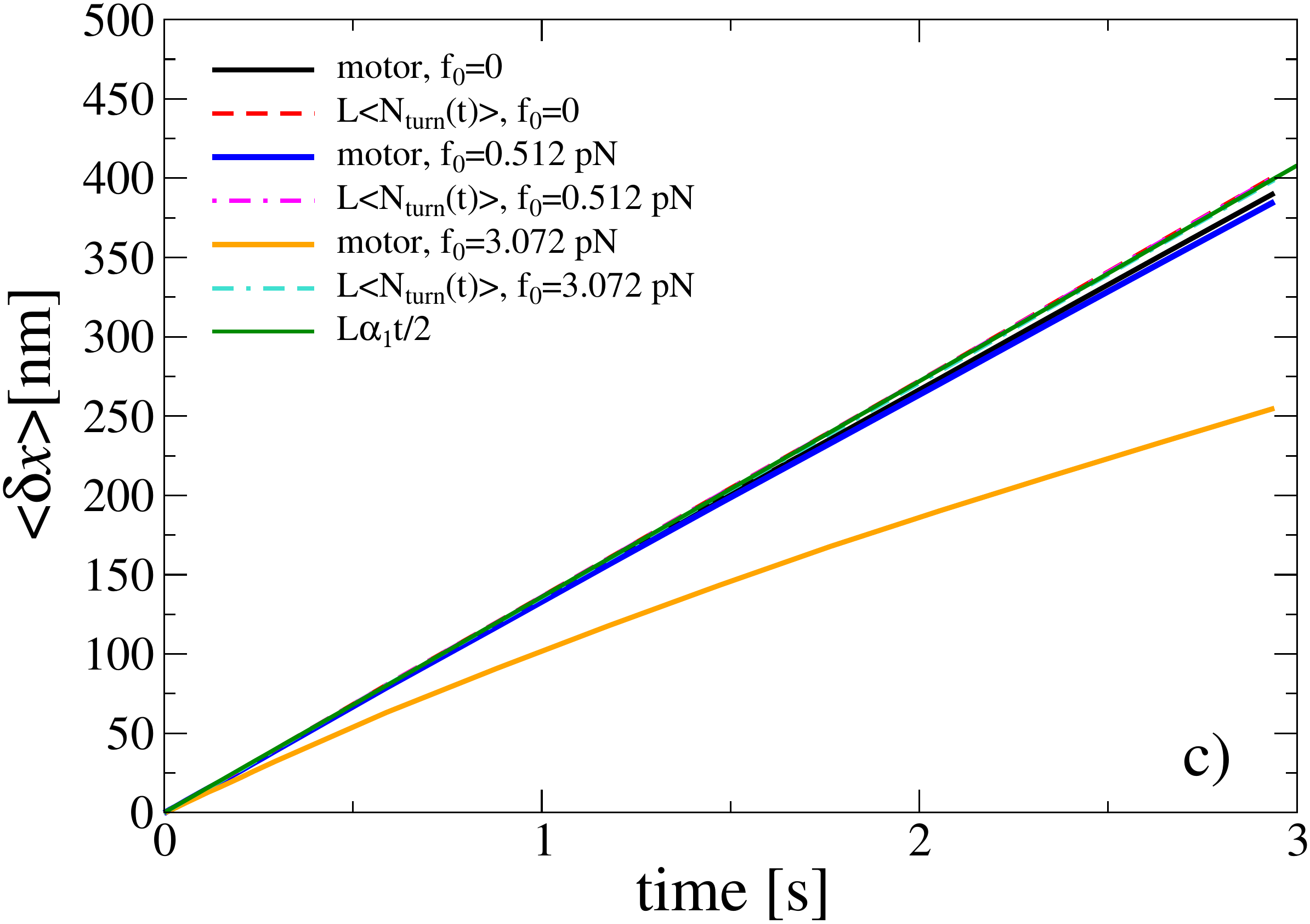}\hspace{0.8cm}
  \includegraphics[width=7cm]{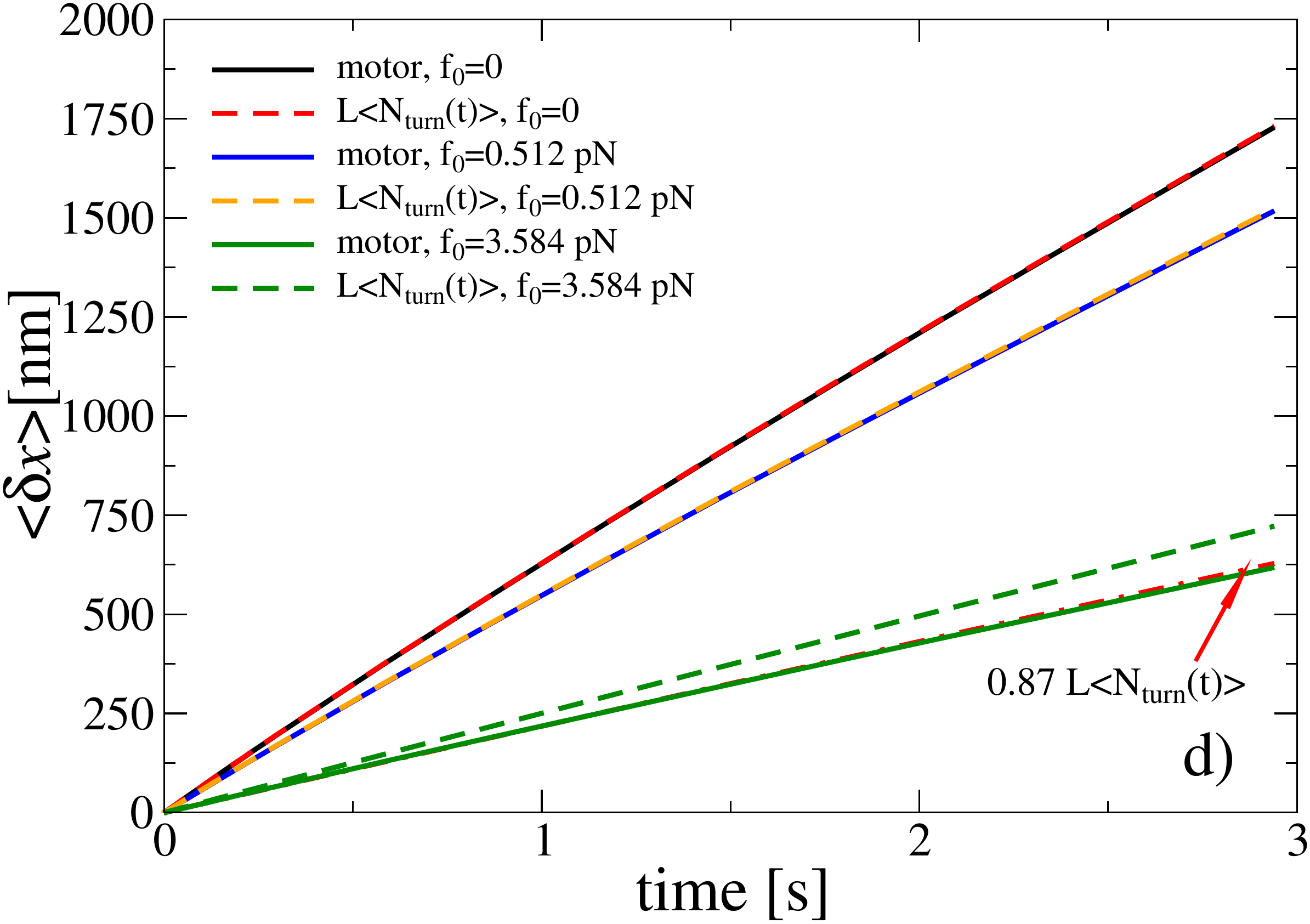} 
 \caption{(color online). Mean motor displacement $\langle \delta x(t)\rangle$ and  $ L\langle N_{\rm turn}(t)\rangle$ \textit{vs.} time for several values of $f_0$ shown in the plots for the sets: (a) $S_2$, model B; (b) $S_6$, model B; (c) $S_5$, model A; (d) $S_5$, model B. Good agreement between $\langle \delta x(t)\rangle$ and $ L\langle N_{\rm turn}(t)\rangle$ reflects a perfect anomalous synchronization between stochastic turnovers of catalytic wheel and the motor stepping along microtubule. For small $f_0$ in the parts (a) and (d), increase of $f_0$  results in a synchronous slowing down of both the biochemical turnovers and the processive mechanical motion. It corresponds to a perfect anomalous ratchet regime. Even at the load corresponding to maximal thermodynamic efficiency in the cases  (a) and (d), only about 15\% and 13\%, correspondingly, of ATP molecules consumed do not result into a perfect transport event -- promotion on the length $L$ along microtubule.  In part (c), $\langle N_{\rm turn}(t)\rangle$ practically does not depend on $f_0$ and is well described by $\alpha_1 t/2$.}
  \label{Fig5}
\end{figure*}

For a smaller cargo, sets $S_2$, the distinction between the models A and B becomes quite evident in Fig. \ref{Fig3}. First, in the model A, $\alpha_{\rm eff}$ starts from $\alpha_{\rm eff}\approx 1$ at small $f_0$, and then it monotonously declines to about $0.8$ at the stalling force. In the model B,
$\alpha_{\rm eff}\approx 0.854$ at $f_0=0$, see in the Table \ref{Table3}. It increases with $f_0$ to about $0.90$ at $f_0$ corresponding to the maximum of thermodynamic efficiency. After this, it declines to about $0.808$ at the stalling force, which is slightly larger than one within the model A. 

\subsubsection{Perfect subdiffusive ratchet}

Within the model A, at small $f_0$ our motor realizes a perfect normal ratchet transport, where stochastic stepping along microtubule is perfectly synchronized with the normal turnovers of the catalytic wheel characterized by a turnover frequency equal to the half of the flashing frequency, \cite{PCCP14,PhysBio15}. A strikingly new result within the model B is that our ratchet realizes a perfect subtransport with anomalous turnovers of catalytic wheel which cannot be characterized anymore by a normal turnover frequency. Rather, one must introduce a new notion, the enzyme catalytic sub-velocity 
$\omega_\gamma$ by $\langle N_{\rm turn}(t)\rangle =\omega_\gamma t^\gamma/\Gamma(1+\gamma)$, see in the Supplementary Material. Notice that an attempt to define the standard turnover rate by $\lim_{t\to \infty}\langle N_{\rm turn}(t)\rangle/t$ would yield zero in this case. As Fig. \ref{Fig5}, a and Table \ref{Table3} reveal, for small $f_0$, $\alpha_{\rm eff}\approx \gamma<1$, and $\lambda\approx 0$. We are dealing with a perfect subdiffusive ratchet, where due to a mechano-chemical coupling, the consumption of ATP molecules by the motor scales sublinearly with time. Nevertheless, the transport is perfect in the sense that consumption of one ATP molecule leads to one step. Indeed, in Fig. \ref{Fig5}, a, $\langle \delta x(t)\rangle$ almost coincides with $L\langle N_{\rm turn}(t)\rangle$ for $f_0=0$,
$f_0=0.512$ pN, $f_0=1.024$ pN. Even for $f_0=4.096$ pN  near to the $R_{\rm th}$ maximum, $\langle \delta x(t)\rangle\approx 0.85 L\langle N_{\rm turn}(t)\rangle$, which means that only about 15\% of biochemical turnovers do not lead to a successful step over $L$. $R_{\rm th}\approx 0.331$ at this maximum is much larger than in the model A, for small cargo, see in Fig. \ref{Fig3}, a. Moreover, this perfect subdiffusive transport is very fast in absolute terms, cf. Fig. \ref{Fig5}, a. Notice, that very differently from the model A, see in Fig. \ref{Fig5}, c, the consumption of ATP molecules strongly depends on $f_0$ within the model B: it is smaller for larger $f_0$ (until about $f_{\rm max}$). This is a very important feature of the perfect subdiffusive ratchet mechanism. It is adaptive and economical.

The transport of the large cargo is far from being perfect in the case $S_1$, model B. However, maybe its quality can be drastically improved at smaller operational frequencies of the motor? Indeed, this is the case, as Fig. \ref{Fig4}, and Fig. \ref{Fig5}, b, reveal for the set $S_6$, model B.  For a smaller $\alpha_2=17\;{\rm s^{-1}}$, $\alpha_{\rm eff}$ increases at $f_0=0$ from about $0.6$ (for $\alpha_2=170\;{\rm s^{-1}}$) to about $0.8$, see in the Table \ref{Table4}, and the maximum of $R_{\rm th}$ increases to about $0.2$, see in Fig. \ref{Fig4}, b, i.e. it almost doubles, cf. Table \ref{Table2}. Even if the quality of anomalous synchronization is somewhat worser in this case than in the case $S_2, B$ of smaller cargo, it is, nevertheless, quite impressive: a heavily loaded motor can walk over the distance of 650 nm at $f_0=0$ (which is normal operational regime of linear molecular motors in living cells) within the less than 3 sec, see in Fig. \ref{Fig5}, b.  Within the model A, transport of large cargo shares similar features for the parameter set $S_5$, with respect to thermodynamic efficiency, see in Fig. \ref{Fig4}, b. However, the dependence of the transport exponent $\alpha_{\rm eff}$ on $f_0$ is entirely different. First, it features an almost normal transport at small $f_0$, cf. Fig.\ref{Fig4}, a, which is a nearly perfect, see in Fig. \ref{Fig5}, c. Second, the turnover frequency of the enzyme practically does not depend on $f_0$, see in Fig. \ref{Fig5}, c. It  equals $\alpha_1/2$. Hence, with the increase of the static load $f_0$ strength, at the maximum of $R_{\rm th}$, $\alpha_{\rm eff}$ drops to about 
$0.81$, see in Table \ref{Table5} and Fig. \ref{Fig4}, while $\gamma$ remains one. This leads to a substantial decay of $R_{\rm th}\propto 1/t^\lambda$ in time, with $\lambda\approx 0.186$. Although, within the model B, set $S_6$, the decay of the maximum of $R_{\rm th}$ has about the same $\lambda\approx 0.196$, see in Table \ref{Table4} at $f_0=3.072$ pN.
This is so because in this case $\gamma$ arrives at the value of one for $f_0=3.072$ pN and larger. Hence, also in this respect, the models A and B are similar. However, once again, the stalling force is slightly larger in the model B.

\subsubsection{The role of the rate $\alpha_2$}

Next, it is interesting to clarify the influence of the rate $\alpha_2$, which is determined, in particular, by the ATP concentration (\cite{AstumianBier,Julicher,Parmeggiani}), on the transport properties within the model B. In fact, the sets $S_2$, $S_5$, and $S_7$ differ only by the value of $\alpha_2$. Fig. \ref{Fig4}, b shows that the smaller is $\alpha_2$, the smaller is the maximum of thermodynamic efficiency, and the smaller is the stalling force. However, at the same time, smaller $\alpha_2$ corresponds to larger $\alpha_{\rm eff}$, see in Fig. \ref{Fig4}, a, i.e. transport becomes closer to normal. Within the model B, the behavior of $\alpha_{\rm eff}$ versus $f_0$ displays one and the same universal feature. First, it slightly increases arriving at a maximum, and then it slightly drops. $\alpha_{\rm eff}$ is generally much less sensitive to $f_0$ within the model B, as compared to the model A. This is because within the model B the mechano-chemical coupling adjusts the tempo of biochemical cycling in response to $f_0$. It becomes slower. Fig.\ref{Fig5}, d and Table \ref{Table6} demonstrate this effect for the set $S_5$, B. Once again, an almost perfect subdiffusive ratchet is realized for a sufficiently small cargo. At $f_0=3.584$ pN, which corresponds to the maximum of $R_{\rm th}$ of about 30\%, only about 13\% of the motor turnovers are futile, not resulting in a successful step along microtubule. A power-stroke like, mechano-chemically adaptive mechanism can  lead to a perfect, energetically efficient subtransport.

\begin{figure}[!]
  \centering
  \includegraphics[width=7cm]{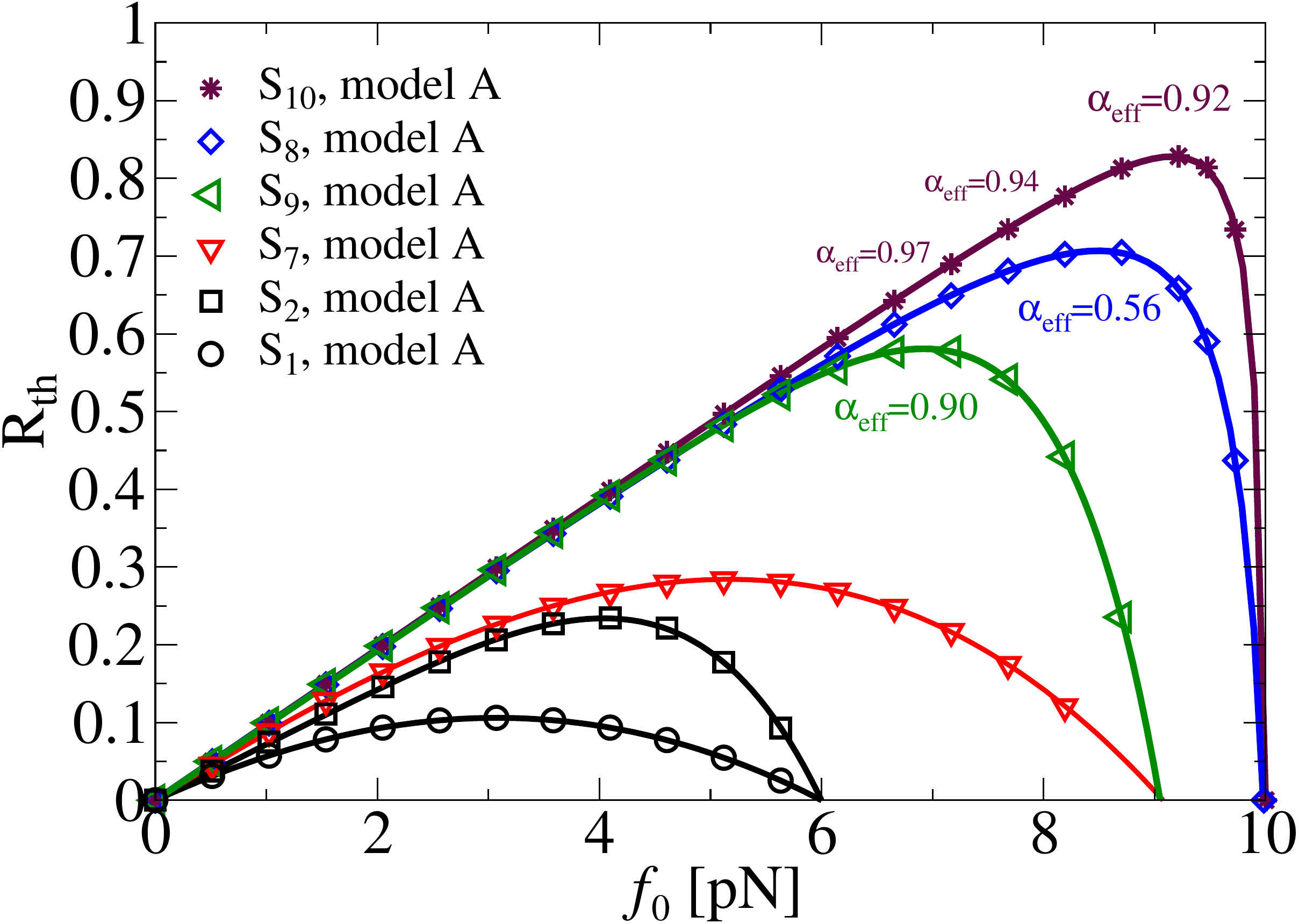}
 \caption{(Color online) Thermodynamic efficiency \textit{ vs.} loading force within model A for $U_0=25\;k_BT_r$ and $U_0=30\;k_BT_r$, as compared with the cases $S_{1,2}$, corresponding to $U_0=20\;k_BT_r$. Full lines present fits with Eq. (\ref{fit}) with parameters shown in Table \ref{Table2}. Notice a substantial increase of efficiency for larger $U_0$. It can exceed 80\% in the case $S_{10}$. Here, numerical data present the results on the proper thermodynamic efficiency as described in the text. It is slightly larger than one in Eq. (\ref{eff2}), see Fig. 6 in \cite{PhysBio15} and the corresponding discussion therein for detail. }
  \label{Fig6}
\end{figure}

\subsubsection{Thermodynamic efficiency over 50\%}

\begin{figure}[!]
  \centering
  \includegraphics[width=7cm]{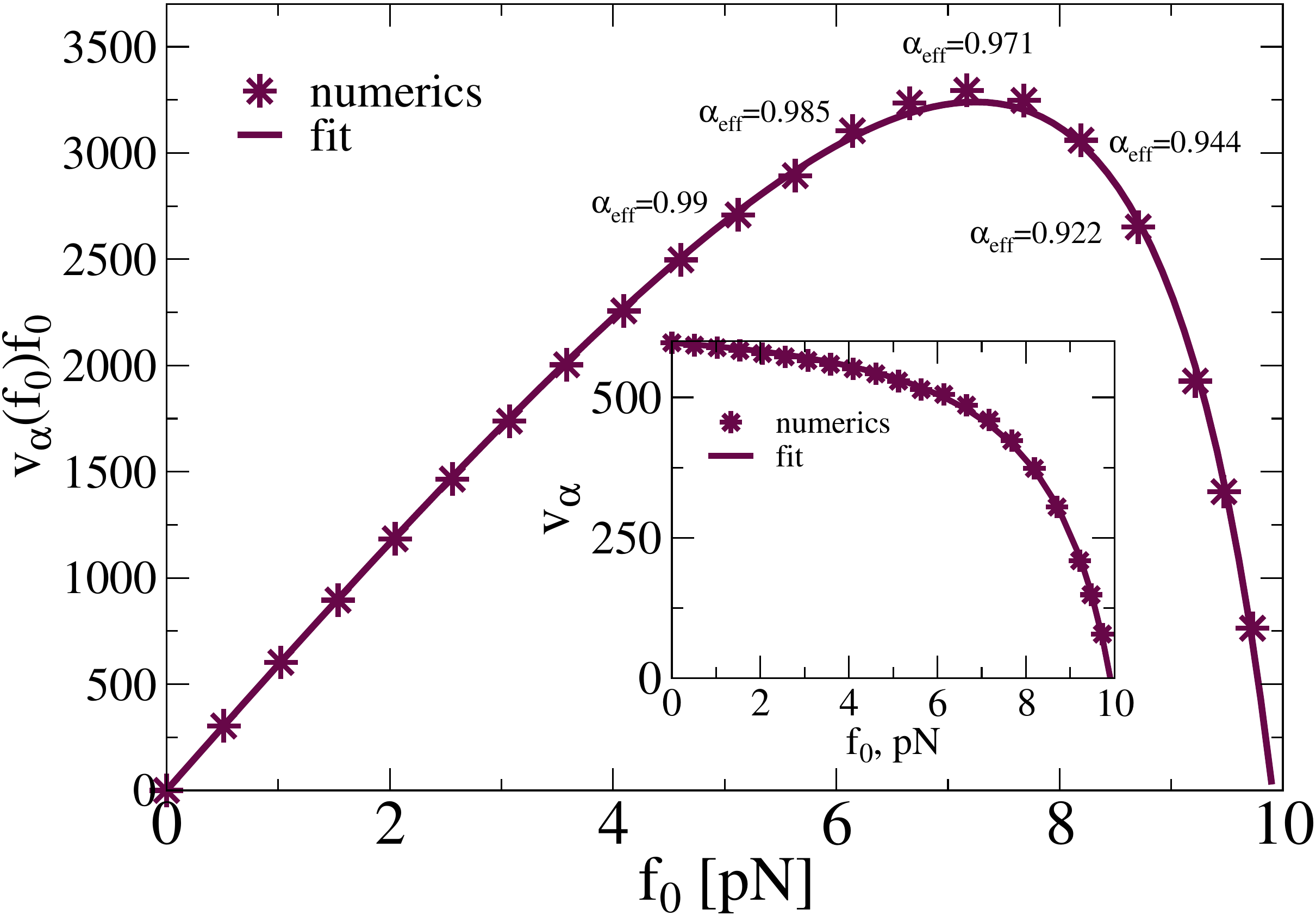}
 \caption{(Color online). Subpower $v_{\alpha}(f_0)f_0$ (in the units of ${\rm pN\cdot nm/s^{\alpha_{\rm eff}}}$) and 
   subvelocity (inset, in the units of ${\rm nm/s^{\alpha_{\rm eff}}}$) versus loading force, in the units of pN, for the set $S_{10}$, model A. $\alpha_{\rm eff}\approx 1$ for $f_0<5$ pN. Several other values are shown in the plot. At the maximum of subpower, at $f_0=7.168$,  $\alpha_{\rm eff}\approx 0.9712$, and $\gamma\approx0.9708$. Hence, the motor operates as a perfect subdiffusive ratchet whose thermodynamic efficiency at the subpower maximum is about 70\%, in accordance with Fig. \ref{Fig6}. The numerical data  are fitted using Eq. (\ref{fit_velo}) with $v_\alpha(0)=597.12\;{\rm nm/s^{\alpha_{\rm eff}}}$, $f_{\rm st}=9.91$ pN, $\epsilon=0.779$, and $q=-7.224$. }
  \label{Fig7}
\end{figure}

Within the model A, the mechano-chemical coupling becomes also very essential, however, for a larger $U_0$. Then, thermodynamic efficiency can overcome 50\%, even at the maximum of sub-power. Figs. \ref{Fig6}, \ref{Fig7} demonstrate this striking effect. For the set $S_{10}$, thermodynamic efficiency exceeds 80\% at its maximum, cf. Fig. \ref{Fig6}. The maximum of $R_{\rm th}$ \textit{vs.} $f_0$ in this case does not corresponds to thermodynamic efficiency at the maximum of sub-power 
$P_{\alpha}(f_0)=v_{\alpha}(f_0)f_0$ because of a strong mechano-chemical coupling. Nevertheless, the maximum of the latter one takes place at $f_0=7.168$ pN in Fig. \ref{Fig7}, which corresponds approximately to impressive 70\% in Fig. \ref{Fig6}. Hence, we provided an instance of anomalous motor whose efficiency at maximal sub-power essentially exceeds 50\%. This is a very important result.
Very interesting is also dependence of the motor (sub)velocity on $f_0$ in this case.
It drops to zero with increasing $f_0$ is a very non-linear fashion, which
is very different from the low efficient quasi-linear regime, where it is nearly linear, \cite{PLoSONE14,PCCP14}. Similar nonlinearities were also observed experimentally for kinesin motors by \cite{Schnitzer00} being, however, fitted in another way following a different model.

\section{Discussion}

The model B exhibits a much stronger mechano-chemical coupling than the model A. Within this model, mechano-chemical coupling is very essential already for $U_0=20\;k_BT_r$, which is a reasonable choice for kinesins II, given a typical stalling force of these motors. While the transport of a large cargo, like magnetosomes in \cite{Robert}, looks very similar in both models, for a fast operating motor, 
the transport of smaller cargos is always profoundly different. The differences are also seen for slowly operating motors. 
Transport of the large cargo in this paper is characterized for $\alpha_2=170\;{\rm s^{-1}}$ by a transport exponent around $\alpha_{\rm eff}=0.6$ for $\alpha=0.4$, which can easily explain the observed superdiffusion exponents around $\beta=1.3\pm 0.1$ in the experiment, \cite{Robert}. Energetically, such a transport is, however, inefficient. Nevertheless, while operating slower the motors can realize also energetically very efficient transport. Indeed, with a tenfold reduction of $\alpha_2$ from 
$170\;{\rm s}^{-1}$ to $17\;{\rm s}^{-1}$, such a near-to-perfect anomalous ratchet regime is realized for $f_0=0$ within the model B (set $S_6$) with $\alpha_{\rm eff}$ increased to about $0.8$. It should be noticed in this respect that normal \textit{modus operandi} of linear motors like kinesin in living cells is one at near-to-zero thermodynamic efficiency. This should not confuse the readers because the useful work is done on overcoming the dissipative resistance of the environment while translocating cargo from one place to another one.
Indeed, the chemical potential of neither motor, not cargo is typically increased. Hence, all the spent energy is eventually dissipated as heat. This is very different from the work of e.g. ionic pumps which must energize ions by transferring them against a corresponding electrochemical gradient. For pumps, namely the thermodynamic efficiency is of paramount importance and it must be optimized. Nevertheless, the ability to sustain substantial constant forces $f_0$ is important for a strong and good motor. It can be checked e.g. in the experiments with optical tweezers. Within the model B, the motor adapts its biochemical cycling to the increased $f_0$. It cycles slower and anomalously, while within the model A  it cycles normally and at the same nearly constant tempo for $U_0=20\;k_BT_r$. 
This advantage of the model B is clearly seen for smaller cargos, where this study revealed a perfect and fast (in absolute terms) anomalous ratchet regime. The motor adapts it cyclic sub-velocity, and even at the maximum of thermodynamic efficiency, while working also against a strong $f_0$, the portion of the futile (in the transport sense) turnovers can be really small, just from 13\% to 15\%. This is definitely provides some benefits with respect to energetic costs of transport.

The mechano-chemical adaptation becomes also relevant  within the model A, however, for larger $U_0$. We showed that for $U_0=0.75$ eV thermodynamic efficiency of our model motor can exceed 80\% within an almost perfect anomalous ratchet regime while transferring smaller cargo against a large bias of $f_0=9$ pN with  $\alpha_{\rm eff}\approx \gamma\approx 0.92$. Also efficiency at maximum sub-power can reach impressive 70\% at $f_0\approx 7.2$ pN with  $\alpha_{\rm eff}\approx \gamma\approx 0.97$.
In such a thermodynamically highly efficient regime, the motor (sub)-velocity declines strongly nonlinearly with $f_0$. Indeed, similar nonlinearities were measured in some experiments with kinesins, \cite{Schnitzer00}. The very existence of such thermodynamically highly efficient regimes is
 especially inspiring when one thinks about perspectives of an optimal motor design, \cite{Cheng,BeilsteinJ}. Clearly, a highly efficient operation is possible also in highly dissipative viscoelastic media like cytosol, as our study convincingly shows. This removes mental barriers and opens great perspectives for an optimal design of artificial molecular motors, \cite{Erbas,Cheng}, especially in allowing to avoid some common fallacy traps, \cite{BeilsteinJ}.

Notice, that within the both studied models of mechano-chemical coupling, we assumed either $\alpha_1$, or $\alpha_2$ be  spatially-independent in a very large (half of the spatial period) region around the potential minima, so that neither $\nu_1(x)$, nor $\nu_2(x)$ turn zero somewhere on microtubule. Allosteric effects are nevertheless present because other rates are spatially dependent.
They can be made stronger, when e.g. $\alpha_1$ in the model A is different from zero only in a small domain around the potential minimum.  Strong allosteric effects are presumably very important for operating natural molecular motors and for designing the new ones, \cite{Cheng}. Such effects can be used for a further optimization of the motor performance, which is very high already in the current, simplified and non-optimized version.

It should be also mentioned that two headed kinesins are highly processive motors, which means that they are attached to microtubule and walk on it before detaching for hundreds of steps and a sufficiently long time of several seconds, \cite{Hancock98}, \cite{Alamilla}. In this respect, the maximal time in our simulations is about 3 sec. Our model is aimed to describe the transport during these processive periods. The influence of viscoelastic environment on their averaged duration, i.e. on the motor processivity, among other factors, \cite{Alamilla}, would also be a very interesting subject for future research, which requires, however, a further generalization of the model considered.\\

\section{Conclusions}

To conclude, in this paper we extended our previous studies of anomalous transport of subdiffusing cargos by molecular motors in viscoelastic cytosol of living cells and showed the emergence of a perfect subdiffusive ratchet regime due to a mechano-chemical coupling. This anomalous transport regime is characterized by anomalously slow biochemical cycling of molecular motors accompanied by a sublinear consumption  of ATP molecules in time, with their optimal use: consumption of one ATP molecule results in one step over the spatial period of microtubule, on average. Moreover, such a transport can be very fast in absolute terms, not bringing some disadvantages in this respect. Such anomalous transport regimes can be very important in the economics of living cells. Their assumed presence provides a true challenge for the experimentalists to reveal. The author hope and expect that the theoretical prediction of a slow consumption of ATP molecules by molecular motors, which cannot be characterized by a standard rate because both the number of enzyme turnovers and the amount of ATP consumed increase sublinearly in time, while transporting efficiently various cargos within interior of living cells, will eventually be confirmed experimentally.

\section*{Acknowledgment} 
Funding of this research by the Deutsche Forschungsgemeinschaft (German Research Foundation), Grant GO 2052/3-1 is gratefully acknowledged.

  \bibliography{biosystems}





\begin{figure*}[ht]
  \centering
  \includegraphics[width=10cm]{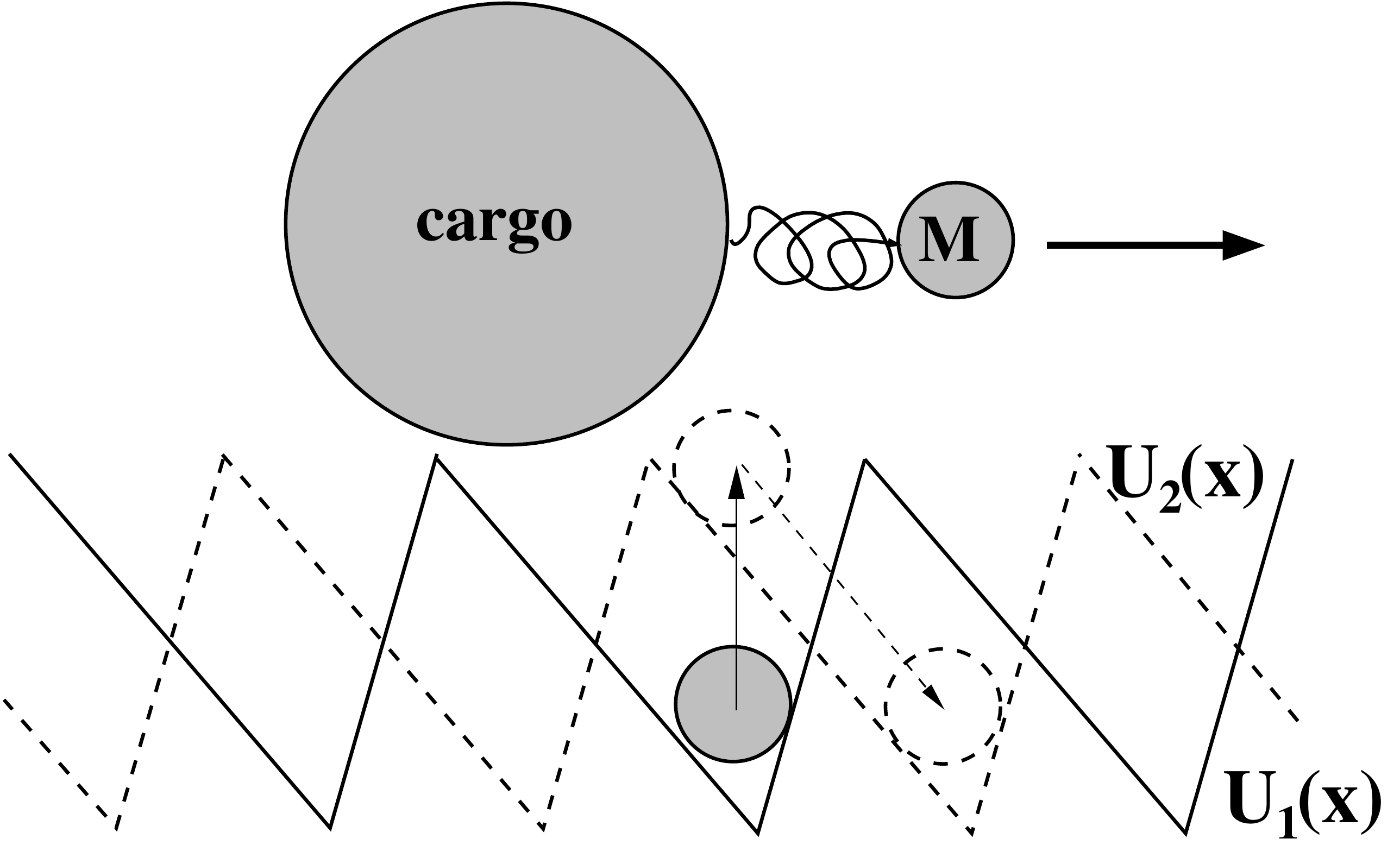}
 \caption{Grafical Abstract}
\end{figure*}

\newpage

\appendix
\newpage
\section{Supplementary Material} 

\subsection{Numerical algorithm}

In this section, a sketch of the numerical algorithm is presented. First, we rewrite Eqs. (7), (11), (12) of the main text as

 \begin{eqnarray}
\label{model}
\dot{x} &=& F(x,\zeta(t))
 +\nu_m G(x,y)+\sqrt{2D_m}\tilde \xi_m(t), \\
\dot{y}&=& -\nu_c G(x,y)-K(y,y_i) + \sqrt{2D_c}\xi_0(t), \nonumber\\
\dot{y_i}&=&\nu_i(y-y_i)+\sqrt{2D_i}\xi_i(t), \nonumber
\label{embedding2}
\label{model1b}
\end{eqnarray}
where $D_m=k_BT/\eta_m$,  $D_c=k_BT/\eta_c$, $D_i=k_BT/\eta_i$ are the corresponding diffusion coefficients, $F(x,\zeta(t))=[f(x,\zeta(t))-f_0]/\eta_m$,
$G(x,y)=(y-x)/[1-(y-x)^2/r_{\rm max}^2]$, $K(y,y_i)=\sum_{i=1}^N\nu_i(y-y_i)$, $\nu_{m}=\kappa_L/\eta_m$, $\nu_{c}=\kappa_L/\eta_c$, and $\tilde\xi_m(t)$ is the scaled $\xi_m(t)$, $\langle \tilde \xi_m(t)\tilde \xi_m(t')\rangle=\delta(t-t')$. Notice that $F(x,\zeta(t))$ can take only two values: $F_-=[-(p+1)U_0/L-f_0]/\eta_m$ or $F_+=[(p+1)U_0/(pL)-f_0]/\eta_m$ depending on the motor position $x$ and conformational state $\zeta(t)=1$, or 
$\zeta(t)=2$. In the state "1": $F(x,1)=F_-$ for $x$ in the interval $[mL,mL+L/(p+1))$, and $F(x,1)=F_+$ for $x$ in the interval $[mL+L/(p+1),(m+1)L)$, where $m$ is an integer number. The values $F(x,2)=F(x+L/2,1)$. The values $F_{\pm}$ alternate in two ways: (i) deterministically depending on the motor position $x$, in the fixed motor state, and (ii) stochastically, when the motor state changes. The latter one is determined as follows.  
To integrate the system of stochastic differential equations (\ref{model}) we use the stochastic Heun algorithm, \cite{GardBook}. This implies that we iterate the time evolution in the discrete times steps $\Delta t$, $t_k=k \Delta t$. Then, for example,
if the motor was in the state "1" at $t_k$, the probability that it makes transition into the state "2" during $\Delta t$ is $p_1=\nu_1(x_k)\Delta t\ll 1$. Hence, one generates a random number $r$ from a uniform distribution on $[0,1]$. If $r\leq p_1$, the transition is done, and otherwise not. Typically, many iterations are required until a transition occurs. Similarly, for the current state "2" with $p_2=\nu_2(x_k)\Delta t\ll 1$. Furthermore,
on each integration time step $\Delta t$ one generates anew $N+2$  independent zero-mean Gaussian variables $W_i$, $W_m$ with unit variance, $i=0,1,2...N$ (Mersenne Twister pseudo-random number generator was used for this). Each propagation step in the discretized time dynamics, $x_k=x(k\Delta t)$,
$y_k=y(k\Delta t)$, $y_{i,k}=y_i(k\Delta t)$, from $t_k=k\Delta t$ to $t_{k+1}=t_k+\Delta t$ consists of two substeps. In the first substep,
\begin{eqnarray}
 x_k^{(1)}&=&x_k+\left [F(x_k,\zeta)+\nu_m G(x_k,y_k)\right ]\Delta t \\
 &+&\sqrt{2D_m\Delta t} W_m, \nonumber \\
  y_k^{(1)}&=&y_k-[\nu_c G(x_k,y_k)+K(y_k,y_{i,k})]\Delta t 
  + \sqrt{2D_c\Delta t} W_0, \nonumber \\
  y_{i,k}^{(1)}& = &y_{i,k}+\nu_i(y_k-y_{i,k})\Delta t 
  + \sqrt{2D_i \Delta t} W_i\;. \nonumber
\end{eqnarray}
In the second (final) step,
\begin{eqnarray}
  x_{k+1}&=&x_k+[F(x_k^{(1)},\zeta)+F(x_k,\zeta)\nonumber \\
  &+&\nu_m G(x_k,y_k)+\nu_m G(x_k^{(1)},y_k^{(1)})]\Delta t/2 \nonumber \\ & + &
 \sqrt{2D_m\Delta t} W_m, \\
  y_{k+1}&=&y_k-[\nu_c G(x_k,y_k)+\nu_c G(x_k^{(1)},y_k^{(1)}) \nonumber \\
 &+& K(y_k,y_{i,k})+K(y_k^{(1)},y_{i,k}^{(1)})]\Delta t/2+\sqrt{2D_c\Delta t} W_0, \nonumber \\
  y_{i,k+1}& = &y_{i,k}+\nu_i(y_k+y_k^{(1)}-y_{i,k}-y_{i,k}^{(1)})\Delta t/2\nonumber \\
  & + &\sqrt{2D_i \Delta t} W_i\;.\nonumber
\end{eqnarray}
Notice that $W_i$ and $W_m$ must be the same numbers on the both substeps within each cycle of iterations, \cite{GardBook}. 
The initial values $x_0=y_0=0$, at the minimum of potential $U_1(x)$ fixed initially, whereas $y_{0,i}$ are sampled from the corresponding Gaussian distributions, see the main text. They are different for each motor particle. The algorithm was implemented in CUDA and propagated in parallel (many different particles with different initially random preparations at the same time) on GPU processors.

\subsection{Fitting dependencies}

A biophysically inspired fitting form reads
\begin{eqnarray}
v_{\alpha}(f)= v_{\alpha}(0) \left [ 1 -\frac{f}{A+B \exp(-\delta f/(k_BT)) } \right ],
\end{eqnarray}
where $A,B$ are some constant with physical dimension of force and $\delta$ is some length. Such dependencies are common in biophysics, \cite{PhillipsBook}. Using the condition $v_{\alpha}(f_{\rm st})$,
this expression can be readily expressed as Eq. (15) of the main text,  with $q=A/f_{\rm st}$ and 
$\epsilon =\delta f_{\rm st}/(k_BT)$.

\begin{figure}[ht]
  \centering
  \includegraphics[width=7cm]{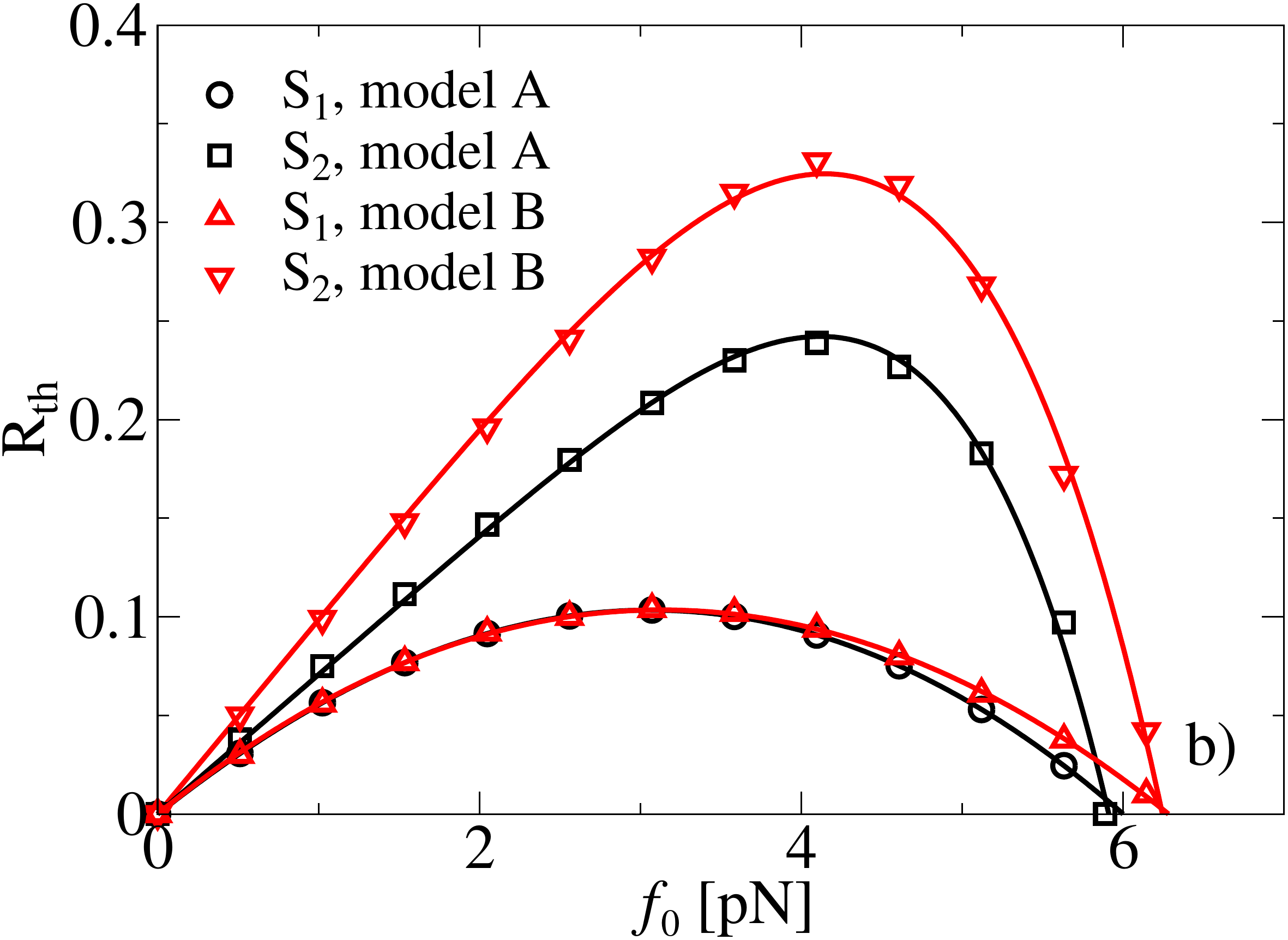} 
 \caption{Thermodynamic efficiency for the sets $S_{1,2}$ in the models A and B \textit{vs.} loading force $f_0$. Full lines present fits with Eq. (\ref{fit}) with parameters shown in Table \ref{Table2}. }
  \label{Fig3bs}
\end{figure}

\begin{figure}[ht]
  \centering
  \includegraphics[width=7cm]{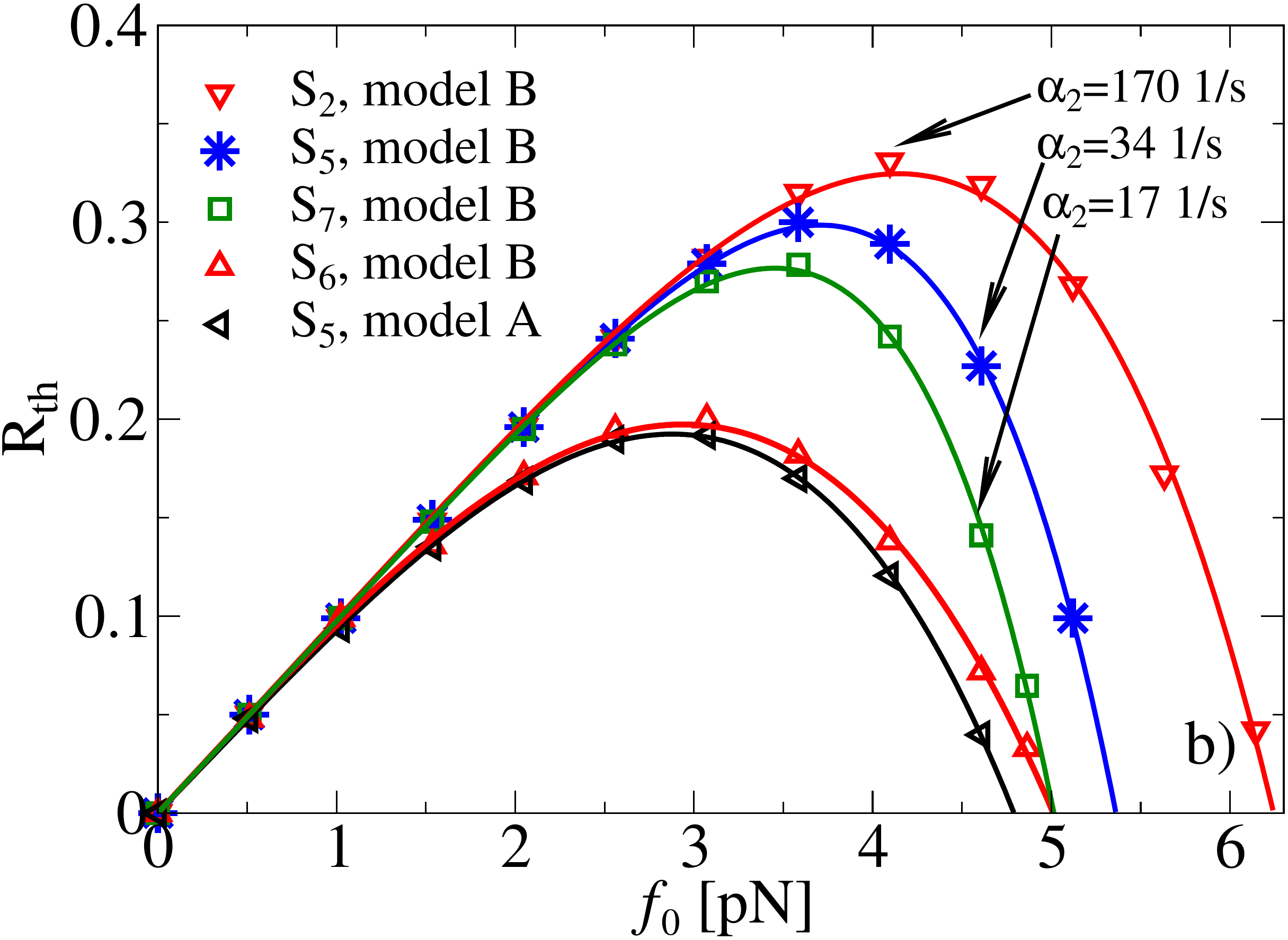} 
 \caption{Thermodynamic efficiency \textit{vs.} loading force $f_0$ for 
 several other sets shown in the plots and discussed in the main text. Full lines present fits with Eq. (\ref{fit}) and parameters shown in Table \ref{Table2}. }
  \label{Fig4bs}
\end{figure}

\begin{figure}[!]
  \centering
  \includegraphics[width=7cm]{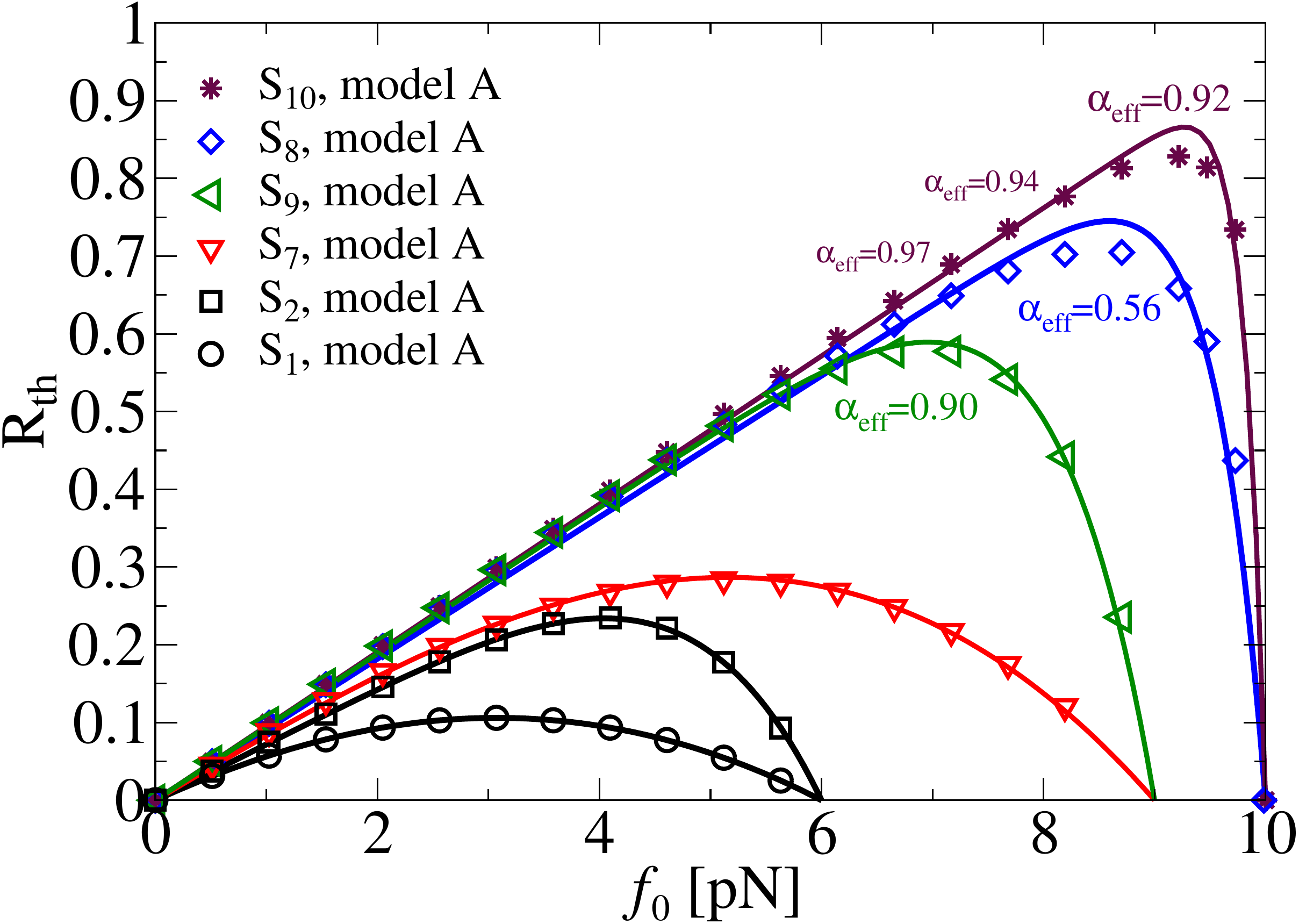}
 \caption{Thermodynamic efficiency \textit{ vs.} loading force within model A for $U_0=25\;k_BT_r$ and $U_0=30\;k_BT_r$, as compared with the cases $S_{1,2}$, corresponding to $U_0=20\;k_BT_r$. Full lines present fits with Eq. (\ref{fit}) with parameters shown in Table \ref{Table2}. }
  \label{Fig6s}
\end{figure}

\begin{figure}[!]
  \centering
  \includegraphics[width=7cm]{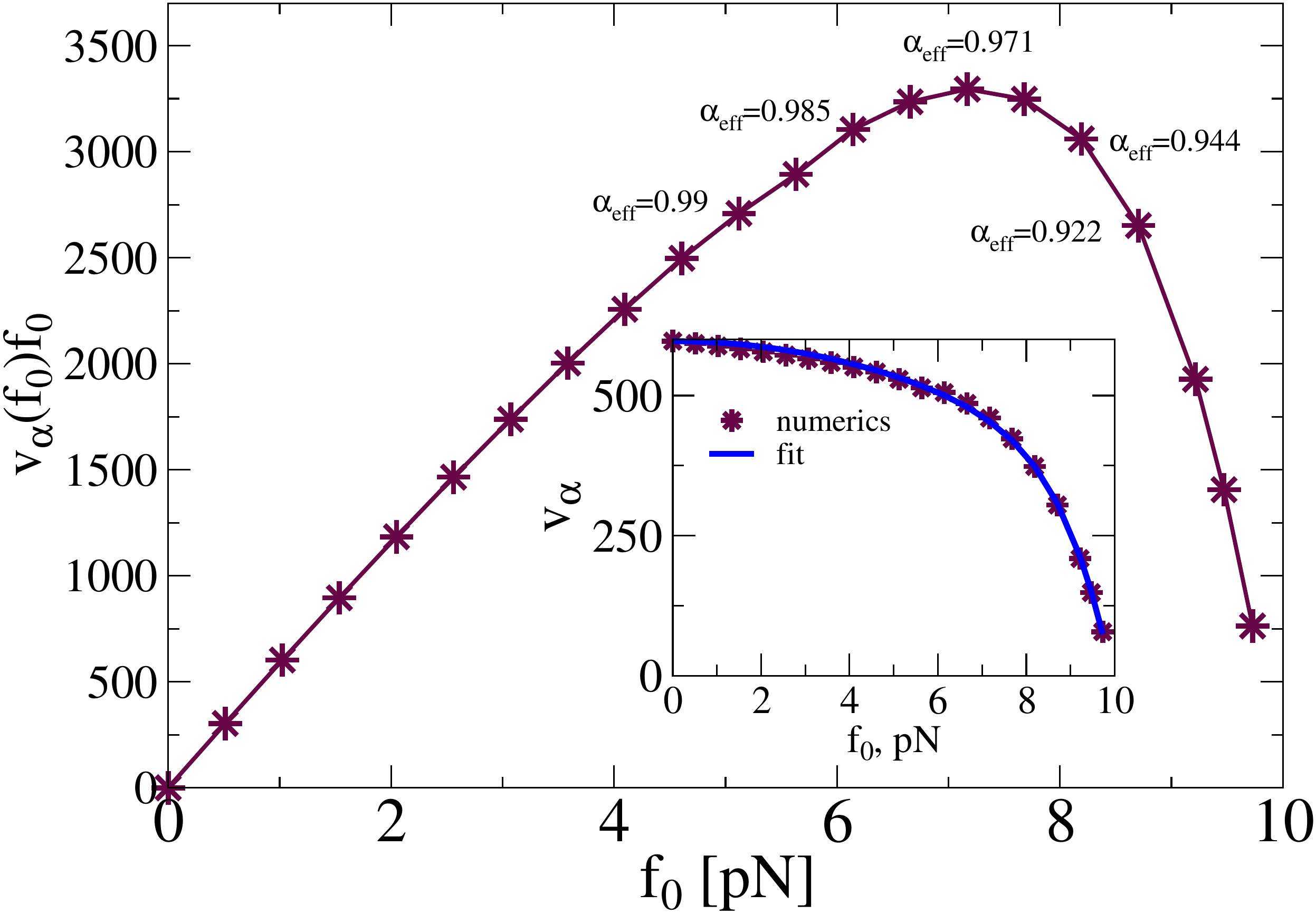}
 \caption{(Color online). Subpower $v_{\alpha}(f_0)f_0$ (in the units of ${\rm pN\cdot nm/s^{\alpha_{\rm eff}}}$) and 
   subvelocity (inset, in the units of ${\rm nm/s^{\alpha_{\rm eff}}}$) versus loading force, in the units of pN, for the set $S_{10}$, model A. The numerical data in inset are fitted by the dependence $v_{\alpha}(f_0)=v_{1}(0)\left [1-(f_0/f_1)^2-(f_0/f_2)^{a_2} \right]$, with $v_1(0)=597.12$ nm/s, $f_1=15.60$ pN, $f_2=10.52$ pN, and $a_2=9.258$.  Fit with Eq. (\ref{another_fit}) (not shown) is essentially  worser. }
  \label{Fig7s}
\end{figure}

Another reasonable form is, \cite{BeilsteinJ},
\begin{eqnarray}\label{another_fit}
v_{\alpha}(f_0)=v_{\alpha}(0)[1-(f_0/f_{\rm st})^a]
\end{eqnarray}
with some fitting power exponent $a$. It has one parameter less. However, a possible interpretation of $a$ is not clear. Then, Eq. (14) is replaced by
\begin{eqnarray}\label{fit}
R_{\rm th}(f_0)=k\frac{f_0}{f_{\rm st}}\left [ 1-\left (\frac{f_0}{f_{\rm st}}\right )^a\right ]. 
\end{eqnarray}
A certain advantage is that the maximal value of $R_{\rm th}^{(\rm max)}=ka/(1+a)^{1+1/a}$ at 
$f_{\rm max}=f_{\rm st}/(1+a)^{1/a}$ can be readily found in analytical form. The corresponding fits
 with the parameters in the Table \ref{Table2} are shown in Figs. \ref{Fig3bs}, \ref{Fig4bs}, \ref{Fig6s}, which correspond to Figs. 3b, 4b, 6, of the main text, respectively. One can see that this alternative fit is really not bad. However, in Fig. 6 of the main text the fitting of two upper curves with Eq. (14) is much better, on the cost of having one parameter more. Actually with $q=0$ in Eq. (14), having the same number of fitting parameters, the both fits are equally good (not shown). However, interpretation of the power exponent $a$, which can take values as large as $52.26$, see in Table \ref{Table2s}, is rather dim. Notice that the inset of Fig. \ref{Fig7s}, which corresponds to Fig. 7 of the main text, contains yet another fit related to (\ref{another_fit}).
 
  \begin{table} 
\begin{center}
\caption{Parameters of the fit $R_{\rm th}(f_0)=k(f_0/f_{\rm st})\left [ 1-(f_0/f_{\rm st})^a\right ]$, and the corresponding values of $R_{\rm th}^{(\rm max)}$, and $f_{\rm max}$}
\label{Table2s}\vspace{0.5cm}

\begin{tabular}{|p{1 cm}|p{1.0 cm}|p{1cm}| p{0.8cm}|p{0.8 cm}|p{1.3 cm}|}
\hline
Set, Model & $k$ & $f_{\rm st}$, pN & $a$ & $R_{\rm th}^{(\rm max)}$ &  $f_{\rm max}$,pN  \\
\hline
$S_{1}, A$ &  0.385 & 6.00 & 1.112 &  0.103 & 3.06\\ 

$S_2$, A &  0.418 & 5.91 & 4.925 & 0.242 & 4.12\\

$S_5$, A &  0.454 & 4.79 & 2.405 & 0.192 & 2.88 \\

$S_7$, A &  0.758 & 9.00 & 1.94 & 0.287 &  5.16\\
$S_8$, A&  0.909 & 10.00 & 20.08 & 0.744 &  8.59 \\
$S_9$, A & 0.847 & 9.00 & 8.93 & 0.589 & 6.96 \\
$S_{10}$, A & 0.952 & 10.00 & 52.26 & 0.866 & 9.27 \\
\hline
$S_{1}$, B &  0.4193 & 6.282 & 0.982 & 0.104 &  3.13\\ 

$S_2$, B &  0.615 & 6.24  &  3.865 & 0.324 &  4.14\\

$S_5$, B &  0.522 & 5.36 & 4.746 & 0.298 & 3.71 \\

$S_6$, B &  0.493 & 5.01  & 2.148 & 0.197 & 2.94 \\

$S_7$, B &  0.489 & 5.01  & 4.619 & 0.277 & 3.45\\

\hline
\end{tabular}
\end{center}
\end{table}

\subsection{Simplest model for anomalous enzyme dynamics}

One of the major results of this paper is that biochemical cycling of a motor enzyme can become anomalously slow and synchronize with its mechanical motion along microtubule, $\gamma\approx \alpha_{\rm eff}<1$, due to influence of viscoelastic environment via a mechano-chemical coupling. This finding can be rationalized within the following (over)simplified model. Cycling of a motor enzyme in its intrinsic conformational space can be parametrized by an angle variable $\phi$. It occurs on a periodic free-energy landscape biased due to energy released in ATP hydrolysis $G(\phi)=G_0(\phi)-\Delta G_{\rm ATP} \phi/(2\pi)$
(\cite{Nelson,Schnitzer00,Alamilla,BeilsteinJ}),  $G_0(\phi+2\pi)=G_0(\phi)$. In other words, $\Delta G_{\rm ATP}$ produces a driving torque $M_{\rm st}=\Delta G_{\rm ATP}/2\pi$. The mechanical load $f_0$ will produce a counter-acting torque $M_0=f_0 L/2\pi$. Here, we assume a perfect synchronization between the mechanical motion and the enzymatic turnover, \cite{Nelson}. Furthermore, let us assume that the conformational motion is subjected to normal and anomalous frictions and the corresponding noise terms, which are related by the fluctuation-dissipation relation (FDR), 
$\langle \zeta_0(t)\zeta_0(t')\rangle=2k_BT r_0 \delta(t-t')$, $\langle \zeta_\gamma (t) \zeta_\gamma(t')\rangle =k_BT r_\gamma/|t-t'|^\gamma$. Then, 
it can be described by a generalized Langevin equation (GLE)
\begin{eqnarray}
\label{model2}
r_0 \dot \phi &= &-\frac{\partial G_0(\phi)}{\partial \phi}+M_{\rm st}-M_0  +\zeta_0(t) \\  &-& r_\gamma \frac{d^\gamma \phi}{dt^\gamma} 
+\zeta_\gamma(t)\;,\nonumber
\end{eqnarray}
where $\frac{d^\gamma \phi}{dt^\gamma}$ is the Caputo fractional derivative, (\cite{Mainardi97,Mathai17}). The just formulated model presents a fractional conformational dynamics generalization of the simplest model of molecular motors, see e.g. in \cite{BeilsteinJ}. The mechanical stalling force is $f_{\rm st }=2\pi M_{\rm st}/L=\Delta G_{\rm ATP}/L$, within this model. With $\Delta G_{\rm ATP}=20\;k_BT_r=82 \;{\rm pN\cdot nm}$, and $L=8$ nm this yields 
$f_{\rm st }=10.25$ pN, which indeed is slightly larger than the maximal stalling force of 10 pN in the main text for $U_0=30\;k_BT_r$. Furthermore, as shown in \cite{GoychukPRE09,GoychukACP12}, in the case of viscoelastic subdiffusion a \textit{static} spatially periodic potential does not influence asymptotically diffusion and transport. Hence, with $t\to \infty$, the number of enzymatic turnovers grows sublinearly as 
\begin{eqnarray}
\langle N_{\rm turn}(t)\rangle \sim \omega_\gamma t^\gamma/\Gamma(1+\gamma),
\end{eqnarray}
where $\omega_\gamma=(M_{\rm st}-M_0)/r_\gamma$ can be termed the catalytic sub-velocity of enzyme. Because the useful work done against the load is 
$W_{\rm use}(t)\sim f_0\langle \delta x(t)\rangle=f_0 L
\langle \phi(t)\rangle/2\pi\propto t^\gamma$  within this model, the thermodynamic efficiency $R_{\rm th}$ is time-independent, and $v_{\alpha=\gamma}(0)=\omega_\gamma L$ in Eq. (\ref{another_fit}). In the $t\to\infty$ limit, $R_{\rm th}$ is given by Eq. (\ref{fit}) with  $k=1$ and $a=1$. It arrives at the maximum of 50\% at $f_{\rm max}=f_{\rm st}/2$.

Of course, the just outlined simplest model of anomalous enzyme turnovers does not correspond precisely to the model in the main text, in some very important detail. First, it restricts the efficiency at maximal sub-power by 50\% -- the Jacobi bound, and corresponds to a symmetric parabolic $R_{\rm th}$ in Eq. (\ref{fit}) with $k=1$, $a=1$. Second, the motor dynamics in the main text was assumed to be normal, memoryless in the absence of a coupled cargo. So, where the memory terms in Eq. (\ref{model2}), the second line, can come from, in principle? The point is that we have to consider some coupling energy $G(\phi,x)$ instead of $G(\phi)$ and to exclude the dynamics of the $x(t)$ variable. Such a procedure generally leads to a memory friction and the related noise in the $\phi$ dynamics considered alone. This is what is assumed in our ultimately simplified model, which does not contain, however, a theory for $\gamma\approx \alpha_{\rm eff}$. In this respect, it must be noted that nonlinear effects in the case of a spatially periodic but fluctuating $G_0(\phi,t)$ are generally very important, \cite{GoychukACP12}. This is the reason why such an oversimplified model cannot describe, e.g., thermodynamic efficiencies over 50\% at the maximum of sub-power, as found and described in the main text.

\end{document}